\documentclass[aps,pre,twocolumn,superscriptaddress,floatfix,showpacs]{revtex4}

\usepackage{amssymb,amsmath,graphicx,bbm}

\begin{document}

\title{Control of amplitude chimeras by time delay in dynamical networks}

\author{Aleksandar Gjurchinovski}
\email{aleksandar.gjurchinovski@gmail.com}
\affiliation{Institute of Physics, Faculty of Natural Sciences and Mathematics, Sts.\ Cyril and Methodius University,
P.\ O.\ Box 162, 1000 Skopje, Macedonia}

\author{Eckehard Sch\"oll}
\email{schoell@physik.tu-berlin.de}

\author{Anna Zakharova}
\email{anna.zakharova@tu-berlin.de}

\affiliation{Institut f\"ur Theoretische Physik, Technische Universit\"at Berlin, 10623 Berlin, Germany}

\begin{abstract}
We investigate the influence of time-delayed coupling in a ring network of non-locally coupled Stuart-Landau oscillators upon chimera states, i.e., space-time patterns with coexisting partially coherent and partially incoherent domains. We focus on amplitude chimeras
which exhibit incoherent behavior with respect to the amplitude rather than the phase and are transient patterns, and show that their lifetime can be significantly enhanced by coupling delay. To characterize their transition to phase-lag synchronization (coherent traveling waves) and other coherent structures, we generalize the Kuramoto order parameter.  
Contrasting the results for instantaneous coupling with those for constant coupling delay, for time-varying delay, and for distributed-delay coupling, we demonstrate that
the lifetime of amplitude chimera states and related partially incoherent states can be controlled, i.e., deliberately reduced or increased, depending upon the type of coupling delay.

\end{abstract}


\date{\today}

\maketitle

\section{Introduction}
A chimera state is a non-trivial partial synchronization state arising spontaneously in networks of identical oscillators \cite{PAN15,SCH16b}. Its dynamics is characterized by a spatio-temporal pattern that consists of two or more spatially separated domains of synchronous (coherent) and asynchronous (incoherent) behavior, respectively. Chimera states have first been reported theoretically in a ring of phase oscillators with symmetric non-local coupling, where for special initial conditions the spatio-temporal domains of in-phase synchronized units coexist with desynchronized domains exhibiting spatially incoherent chaotic dynamics \cite{KUR02a,ABR04}. This observation was unexpected, because the chimera state occurs for the same parameters at which the fully coherent state is stable. The discovery of this intriguing bistable phenomenon initiated a large number of theoretical investigations of chimera states, including their bifurcation properties \cite{ABR08}, their occurrence in various network topologies \cite{MAR10a,KUR03,MAR10b,OME12,PAN13,PAN15a}, and their emergence for a wide range of local oscillatory dynamics \cite{SCH16b}.

The bistability of chimera states and the fully synchronized solution, and the smallness of the basin of attraction of the chimera states was the main cause for detecting chimera states in real experiments only a decade after their discovery in numerical simulations \cite{HAG12,TIN12,NKO13}. Chimera states may as well exist outside of carefully prepared laboratory setups, and are believed to be related to the unihemispheric sleep of various types of mammals, birds \cite{RAT16} and humans \cite{TAM16}, to epileptic seizure \cite{ROT14}, power grid blackouts \cite{MOT13}, social networks \cite{GON14}, and various neural networks. 

Of particular practical importance is the question of stability of chimera states under different types of perturbations, e.g., noise, and of their lifetime: are the chimeras only transient states or are they stable persisting phenomena, and under what conditions can their lifetime be enhanced or reduced? 
It has been shown theoretically for phase oscillators in non-locally coupled ring networks that chimera states are transients, and their lifetime increases exponentially with the system size as the number of oscillators increases \cite{WOL11,WOL11a,OME13a}, similar to transient chaos in spatially extended systems \cite{WAC95b,LAI11a,TEL15}. Controlling the lifetime and basin of attraction of chimera states is therefore crucial in practical application, and some progress in this direction has been made only recently 
\cite{SIE14c,LOO16,OME16}. It is well known that generally time-delayed feedback or coupling is a powerful method to control the stability in nonlinear systems \cite{SCH07} and networks \cite{SCH13}.   
 
In this paper, we investigate the influence of time delay on chimera states in delay-coupled oscillator networks, specifically, the dynamical regimes and the lifetime of amplitude chimeras in a ring network of Stuart-Landau oscillators with nonlocal time-delayed coupling. Amplitude chimeras occur in networks with coupled phase and amplitude dynamics, and they are distinguished from phase chimeras and amplitude-mediated chimeras \cite{SET14} by the fact that coherence and incoherence occur only with respect to the amplitude of the oscillators while all the elements of the network oscillate periodically with the same frequency and correlated phase \cite{ZAK14},
and they have been found in Stuart-Landau networks with coupling which breaks the rotational $S^1$ symmetry. In contrast to classical chimeras \cite{WOL11} for amplitude chimeras the transient time decreases and saturates for large system size \cite{LOO16}.

In Section II we introduce the model equations and describe the characteristics of the occurring amplitude chimera states and the related chimera death steady states, which have recently been discovered \cite{ZAK14,ZAK15b}. In Section III we generalize the global Kuramoto order parameter as a measure to characterize the transient time of chimera states towards different types of completely coherent states, in particular, phase-lag synchronization (traveling wave), and other coherent states with more complicated waveform structures. In Section IV a detailed analysis of the dynamical regimes and the transient time of amplitude chimeras is provided in the case of (i) no delay in the coupling, (ii) constant delay, (iii) deterministic time-varying delay, and (iv) distributed delay in the coupling. The conclusions are given in Section V.

\section{The model}

We investigate non-locally coupled ring networks of $N$ oscillators with different types of delay
in the coupling introduced via the delay operator $\mathcal{D}(\cdot)$. The local dynamics of the nodes is described by the Stuart-Landau oscillator, i.e., the normal form of a supercritical Hopf bifurcation \cite{KUR02a,ATA03,CHO09,KYR13,SCH13b,POS13a,ILL16,PAN16a}. The dynamical equations are given by:
\begin{align}
\dot z_j=&\left[\lambda+i\omega-|z_j|^2\right]z_j\nonumber\\
&+\frac{\sigma}{2P} \sum_{k=j-P}^{j+P}\left[\mathrm{Re}(\mathcal{D}[z_k(t)])-\mathrm{Re}(z_j(t))\right],
\label{eq_sys}
\end{align}
where $z_j=x_j+iy_j \in \mathbbm{C}$, $\lambda, \omega \in \mathbbm{R}$. The variable $z_j$ describes the $j$-th node ($j=1,2\dots N$, all indices mod $N$), $\sigma$
is the coupling strength, and $P$ is the number of coupled neighbors in each direction on a ring. In polar coordinates, $z_j=r_j e^{i\theta_j}$, where $r_j=|z_j|$ and $\theta_j=\mathrm{arg}(z_j)$, the dynamics of the uncoupled system is given by $ \dot r_j=(\lambda- r_j^2)r_j$ and $ \dot\theta_j=\omega$. For $\lambda>0$, the uncoupled delay-free system exhibits self-sustained limit cycle oscillations with radius $r_0=\sqrt{\lambda}$ and frequency $\omega$. Since in our simulations we fix $\omega=2$, the period of oscillations is $T=\pi$, which we will also refer to as intrinsic period. Depending on the type of the delay in the coupling, the delay operator $\mathcal{D}$ could act upon the state function $z(t)$ in different forms, e.g.
$\mathcal{D}_1[z(t)]=z(t-\tau)$
in the case of \emph{constant} delay,
$\mathcal{D}_2[z(t)]=z(t-\tau(t))$ 
for \emph{time-varying} delay, with time dependence given by the function $\tau(t)$, or
$\mathcal{D}_3[z(t)]=\int\limits_{0}^{\infty} G(t')z(t-t')\mathrm{d}t' $
for \emph{distributed} delay, where $G(t)$ is a kernel characterizing the delay distribution.
Here we consider coupling only in the real parts, since this breaks the rotational $S^1$ symmetry of the system which is a necessary condition for the existence of nontrivial steady states $z_j \neq 0$ and thus for oscillation death \cite{ZAK13a}. Therefore, the symmetry-breaking form of the interaction between the oscillators induces a set of inhomogeneous fixed points in addition to the homogeneous fixed point at the origin $r_j=0$.

\begin{figure}
\centering
$\begin{array}{ll}
\includegraphics[width=0.55\columnwidth]{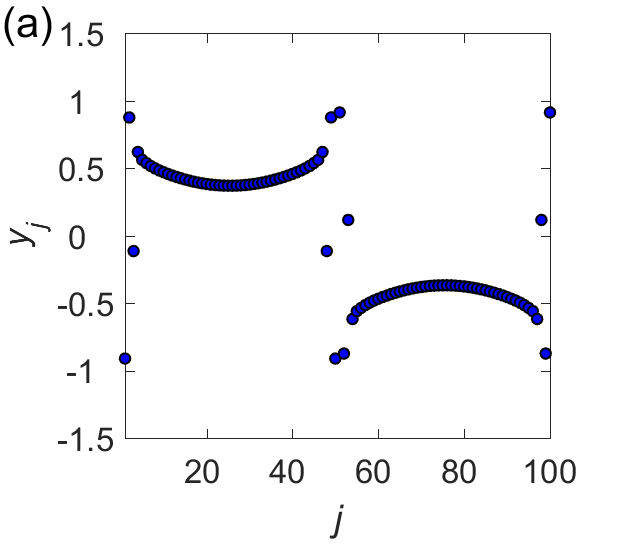}&
\includegraphics[width=0.43\columnwidth]{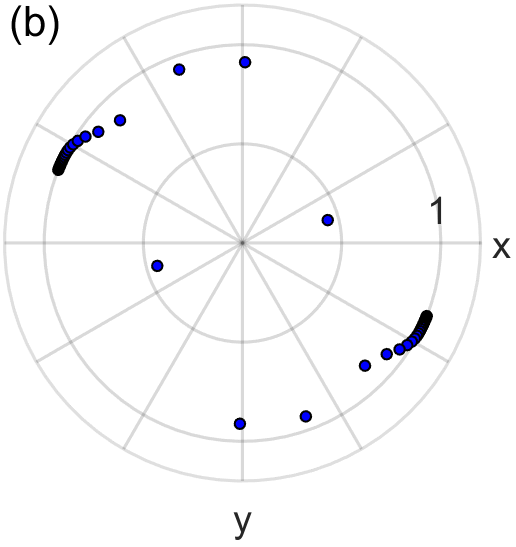}\\
\includegraphics[width=0.55\columnwidth]{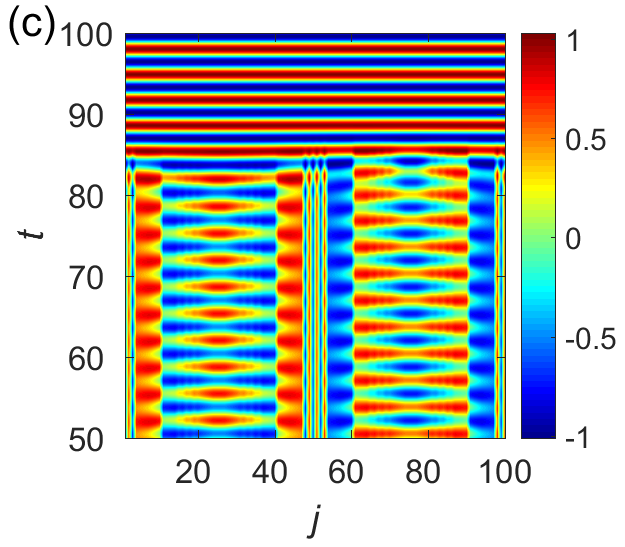}&
\includegraphics[width=0.43\columnwidth]{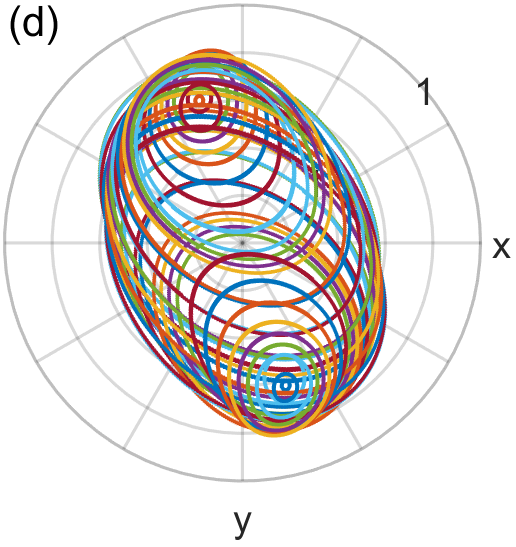}\\
\includegraphics[width=0.55\columnwidth]{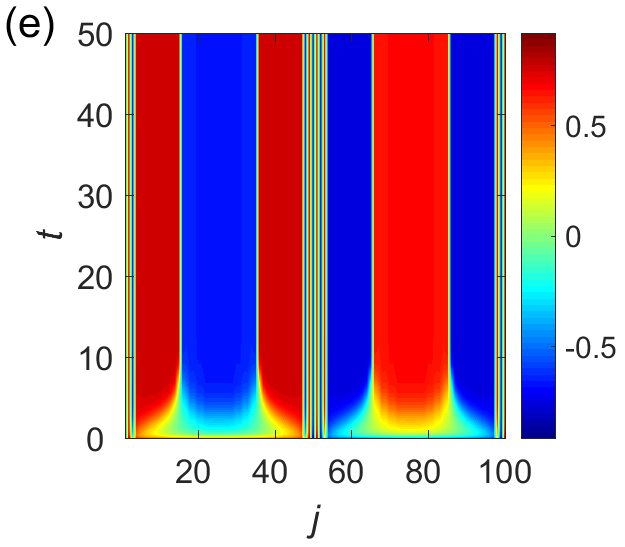}&
\includegraphics[width=0.43\columnwidth]{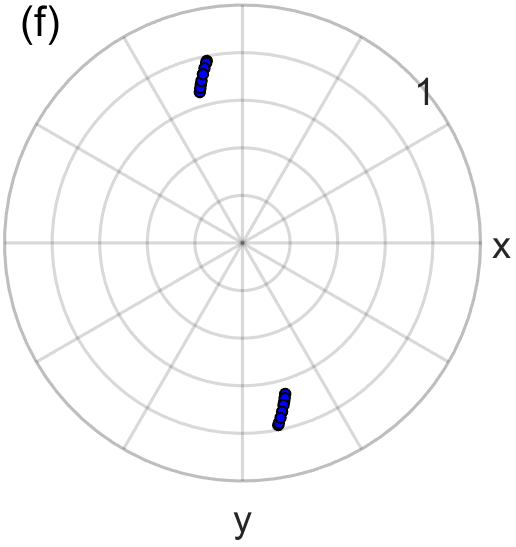}
\end{array}$
\caption{(Color online) (a,b) Initial condition ($t \le 0$) used in the simulations: Snapshot of amplitude chimera state $y_j=\mathrm{Im}(z_j)$ (panel a) and phase portrait in the complex $z$-plane (panel b) for $P=1$, $\sigma=14$, $\tau=0$. 
(c,d) Space-time plot of $y_j=\mathrm{Im}(z_j)$ showing the collapse of initial amplitude chimera towards in-phase synchronized regime at $t\approx 84$ (panel c) and phase portrait of all oscillators for the time window $t\in[60,65]$ (panel d) for $P=16$, $\sigma=7$, $\tau=0$. 
(e,f) Space-time plot of $y_j=\mathrm{Im}(z_j)$ showing the collapse of initial amplitude chimera towards 3-cluster chimera death state (panel e) and phase portrait of all oscillators for the time window $t\in[60,90]$ (panel f) for $P=31$, $\sigma=9$, $\tau=0$. 
Other parameters: $\lambda= 1$, $\omega = 2$, $N=100$.}
\label{fig4_5}
\end{figure}

In the instantaneous coupling case $\mathcal{D}_0[z(t)]=z(t)$, the system Eq.(\ref{eq_sys}) exhibits various dynamical regimes due to the interplay of the symmetry-breaking coupling between the individual oscillators and the nonlocal network topology, e.g.,
multi-cluster oscillation death \cite{SCH15b}.
In particular, this system demonstrates chimera behavior with respect to amplitude dynamics, i.e., amplitude chimeras \cite{ZAK14,ZAK15b,ZAK16}, where one part of the network is oscillating with spatially coherent amplitude, while the other displays oscillations with spatially incoherent amplitudes and centers of mass (Figs. \ref{fig4_5}a-d). 
For all parameter values we choose the same initial condition (Figs. \ref{fig4_5}a,b) which is an amplitude chimera for $P=1$, $\sigma=14$, and is constructed by simulating the system starting from a fully antisymmetric state: the first half of the nodes $j \in{(1, ..., N/2)}$ is set to $({x}_{j},{y}_{j}) = (-0.9, -0.9)$, whereas the second half $j \in{(N/2 +1, ..., N)}$ is set to$({x}_{j},{y}_{j}) = (0.9, 0.9)$.
An important feature of this pattern is that within its incoherent domain the center of mass of each oscillator is shifted away from the origin, while the nodes belonging to the coherent population oscillate around the origin and with larger amplitude. This becomes particularly obvious from the phase portrait of amplitude chimera (Fig. \ref{fig4_5}d). Another crucial feature of the amplitude chimera is that averaged phase velocities remain the same for every element of the network, since the phases are correlated even within the incoherent domain. In other words, amplitude chimeras demonstrate chimera behavior exclusively with respect to amplitude dynamics rather than the phase, in contrast to amplitude-mediated chimeras, for which both phase and amplitude are in a chimera state \cite{SCH14a,SET13}. Amplitude chimera states (Fig. \ref{fig4_5}c) can appear as long transients, potentially lasting for hundreds or even thousands of oscillation periods before a coherent state is reached.

Another chimera pattern observed in system Eq.(\ref{eq_sys}) shows chimera behavior of steady states and is called {\it chimera death} \cite{ZAK14,ZAK15b,ZAK16} (Figs. \ref{fig4_5}e,f). In this regime the oscillations are quenched in a peculiar way. The population of identical elements breaks up into two groups: (i) spatially coherent inhomogeneous steady state (oscillation death), where the neighboring nodes of the network are correlated forming a regular inhomogeneous steady state, and (ii) spatially incoherent oscillation death, where the sequence of populated branches of the inhomogeneous steady state of neighboring nodes is completely random (Fig. \ref{fig4_5}e). The term ``coherent/incoherent'' in this case refers to the coherence/incoherence in space, i.e., spatial correlation, which is different from temporal coherence (correlations of the temporal dynamics). Interestingly, the variation of the coupling range for fixed coupling strength in system Eq.(\ref{eq_sys}) leads to the formation of clusters within the coherent domain of chimera death: with increasing coupling range the number of clusters in the coherent spatial domain is decreased. A 3-cluster chimera death is illustrated in Fig. \ref{fig4_5}e.

\section{Characterizing the transition from incoherence to coherence}

The state of the network at a given time can be quantitatively described by introducing a global order parameter \cite{WOL11}, but also other measures have been suggested \cite{KEM16}. 
This is especially important for finding the transient times of different states. In particular, we consider the transition from 
partially incoherent states, such as phase, amplitude or amplitude-mediated chimeras towards various coherent space-time 
structures, such as in-phase synchronization, phase-lag synchronization (traveling waves), and more complicated 
waveforms with different degree of coherence.

\begin{figure*}
\centering
\includegraphics[width=0.245\textwidth]{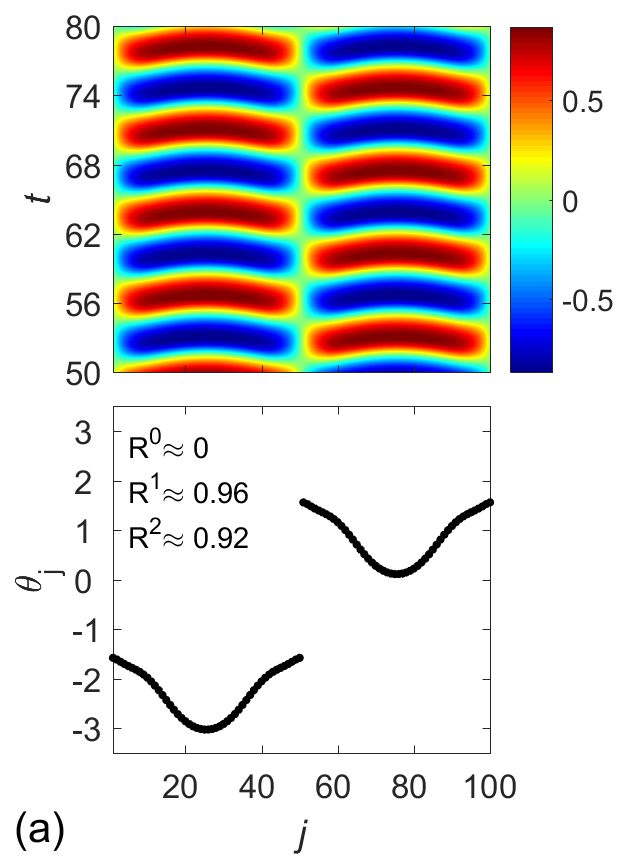}
\includegraphics[width=0.245\textwidth]{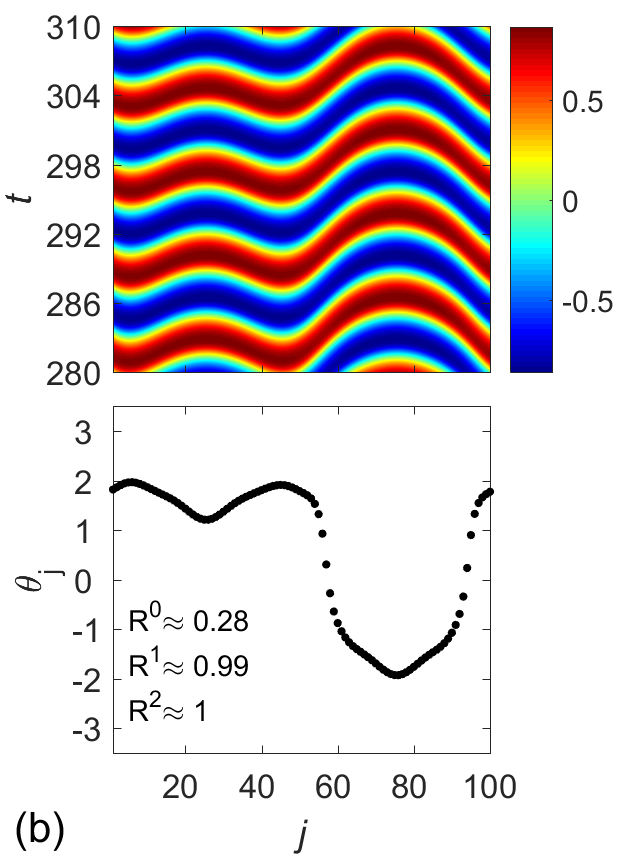}
\includegraphics[width=0.245\textwidth]{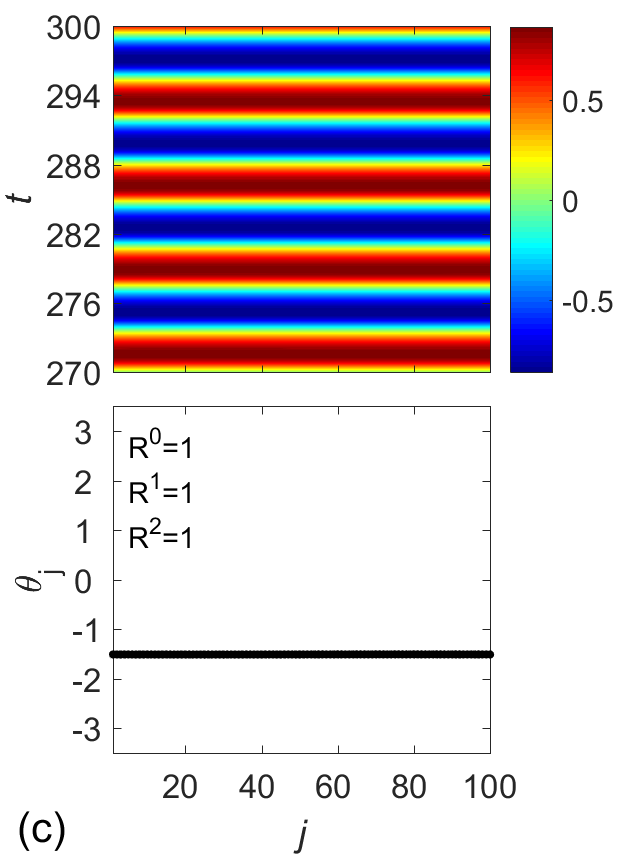}
\includegraphics[width=0.245\textwidth]{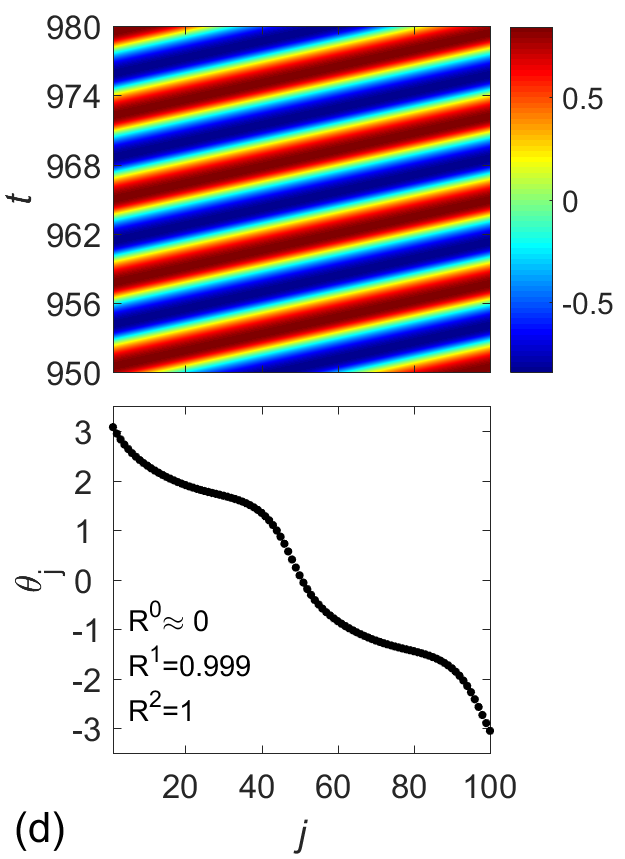}\\
\caption{(Color online) Coherent structures in a ring network of $N=100$ Stuart-Landau oscillators for $\sigma=3$ and constant delay $\tau=\pi/2$. 
Upper panels: space-time plots of the variable $y_j=\mathrm{Im}(z_j)$. 
Lower panels: snapshots of the phase $\theta_j$ at time $t=t_s$. 
(a) anti-phase clustering, $P=7$, $t_s=66$. 
(b) distorted sine-like wave, $P=2$,  $t_s=296.5$. 
(c) zero-lag synchronization, $P=23$, $t_s=275$. 
(d) imperfect phase-lag synchronization (traveling wave), $P=7$, $t_s=959.8$.
Corresponding mean field parameters at $t=t_s$: (a) $R^0 \approx 0$, $R^1=0.96$, $R^2=0.92$; 
(b) $R^0=0.28$, $R^1=0.99$, $R^2=1.00$; (c) $R^0=R^1=R^2=1.00$; (d) $R^0\approx 0$, $R^1=0.999$, $R^2=1.00$.
Other parameters: $\lambda= 1$, $\omega = 2$.}
  \label{fig1}
\end{figure*}

We define a snapshot of the network dynamics as a set of values $\{z_1(t), z_2(t),\dots z_N(t)\}$ of the variables at a fixed time $t$.
By examining the distribution of the state vectors $z_j(t)=|z_j|\exp(i\theta_j)$ in the complex $z$ plane, one can extract the information on the dynamics at a given instant and, in particular, trace the appearance of spatially coherent structures.

First, we adopt the global Kuramoto order parameter 
\begin{equation}
    R^0 = \left|\frac{1}{N}\sum_{j=1}^{N}e^{i\theta_j}\right|,
    \label{kuramoto.order.param}
\end{equation}
as a measure to distinguish between coherent and incoherent states of the network, which we refer to as a \emph{zeroth-order} mean field parameter in the following. For complete in-phase synchronization, the phases $\theta_j$ of all 
oscillators are constant, leading to $R^0=1$; on the other hand, for complete desynchronization $R^0=0$. However, there are other coherent patterns such as anti-phase clustering (standing waves), where two groups of nodes oscillating in anti-phase are separated by elements which are almost stationary.  Such coherent patterns or different types of traveling waves cannot be detected by $R_0$ (see Fig.\ref{fig1}).

In the case of coherent traveling waves, i.e., constant phase-lag synchronization (splay state), the phases $\theta_j$ 
are \emph{not} constant, but the \emph{phase differences} between adjacent oscillators are constant. 
To characterize such a network state, we introduce the \emph{first-order} mean field parameter
\begin{equation}
R^1 = \left|\frac{1}{N}\sum_{j=1}^{N}e^{i\Delta\theta_j}\right|,
\label{g1.formula}
\end{equation}
as an extension of the Kuramoto order parameter $R^0$, 
where the phase differences between adjacent oscillators are used instead of the phases $\theta_j$:
\begin{equation}
\Delta\theta_j = \theta_{j+1} - \theta_{j}.
\label{firstorderdiff}
\end{equation}
Since for the traveling waves $\Delta\theta_j\simeq const$, this regime is characterized by $R^1\approx 1$, while $R^0$ in this case cannot be used as an indicator of coherence ($R^0\approx0$) (see Fig. \ref{fig1}d).

By looking at the snapshots of the phase in the $(\theta_j(t),j)$ plane at a given time $t$, one observes that an in-phase synchronized regime has a constant $\theta$ profile, i.e., a horizontal line (see Fig.\ref{fig1}c).
On the other hand, in the ideal case of no fluctuations of the phase differences, the phase-lag synchronization (traveling wave) is characterized by an inclined line in the ($\theta_j$, $j$) plane,
such that the \emph{first derivative} ${d\theta_j}/{d j}$ is constant (see Fig. \ref{fig1}d for an example of an imperfect traveling wave). 
Strictly speaking, this derivative is taken in the continuum limit $N\rightarrow\infty$.
In our finite difference case, we have
\begin{equation}
\frac{\Delta\theta_j}{\Delta j} = \Delta\theta_j
\end{equation}
since $\Delta j = (j+1) - j=1$, which motivates the definition of $R^1$ in Eq. (\ref{g1.formula}).

These coherence measures could be further generalized to higher orders to describe more complicated waveforms 
(e.g., sine-shaped waves and other distorted waveforms) that may arise after an amplitude chimera collapse (see Fig. \ref{fig1}b).
These distorted waves are also coherent structures, but their appearance is difficult to detect by the previously introduced order parameter.
Therefore, we define the \emph{second-order} mean field parameter:
\begin{equation}
R^2 = \left|\frac{1}{N}\sum_{j=1}^{N}e^{i\Delta^2\theta_j}\right|,
\end{equation}
where now the difference of the phase differences is used, 
\begin{equation}
\Delta^2\theta_j = \Delta\theta_{j+1} - \Delta\theta_{j},
\end{equation}
which, according to Eq. (\ref{firstorderdiff}) can be written as:
\begin{equation}
\Delta^2\theta_j = \theta_{j+2} - \theta_{j+1} - (\theta_{j+1} - \theta_{j}) = \theta_{j+2} - 2\theta_{j+1}+ \theta_{j}.
\end{equation}
The \emph{second-order} mean field parameter 
$R^2=1$  characterizes second order wave-like coherent structures with constant curvature.

For higher order coherent structures one may define the \emph{n-th order} mean field parameter:
\begin{equation}
R^n = \left|\frac{1}{N}\sum \limits_{j=1}^{N}e^{i\Delta^n\theta_j}\right|,
\end{equation}
with the $n$-th order forward difference:
\begin{equation}
\Delta^n\theta_j = \sum \limits_{k=0}^n (-1)^k \binom{n}{k} \theta_{j+n-k}.
\end{equation}

The mean-field parameters $R^n$ defined above can be used to describe various coherent states of the network independently of its topology and the fact whether the nodes are identical or not. The newly introduced order parameters as a natural extension of the global Kuramoto order parameter are complementary measures which allow us to discriminate between different transient patterns and to determine their lifetime. Their application is not confined to amplitude dynamics, and they can also be used to describe 
various transient phenomena in the case of phase oscillator networks. 

\begin{figure*}
\centering
\includegraphics[width=\textwidth]{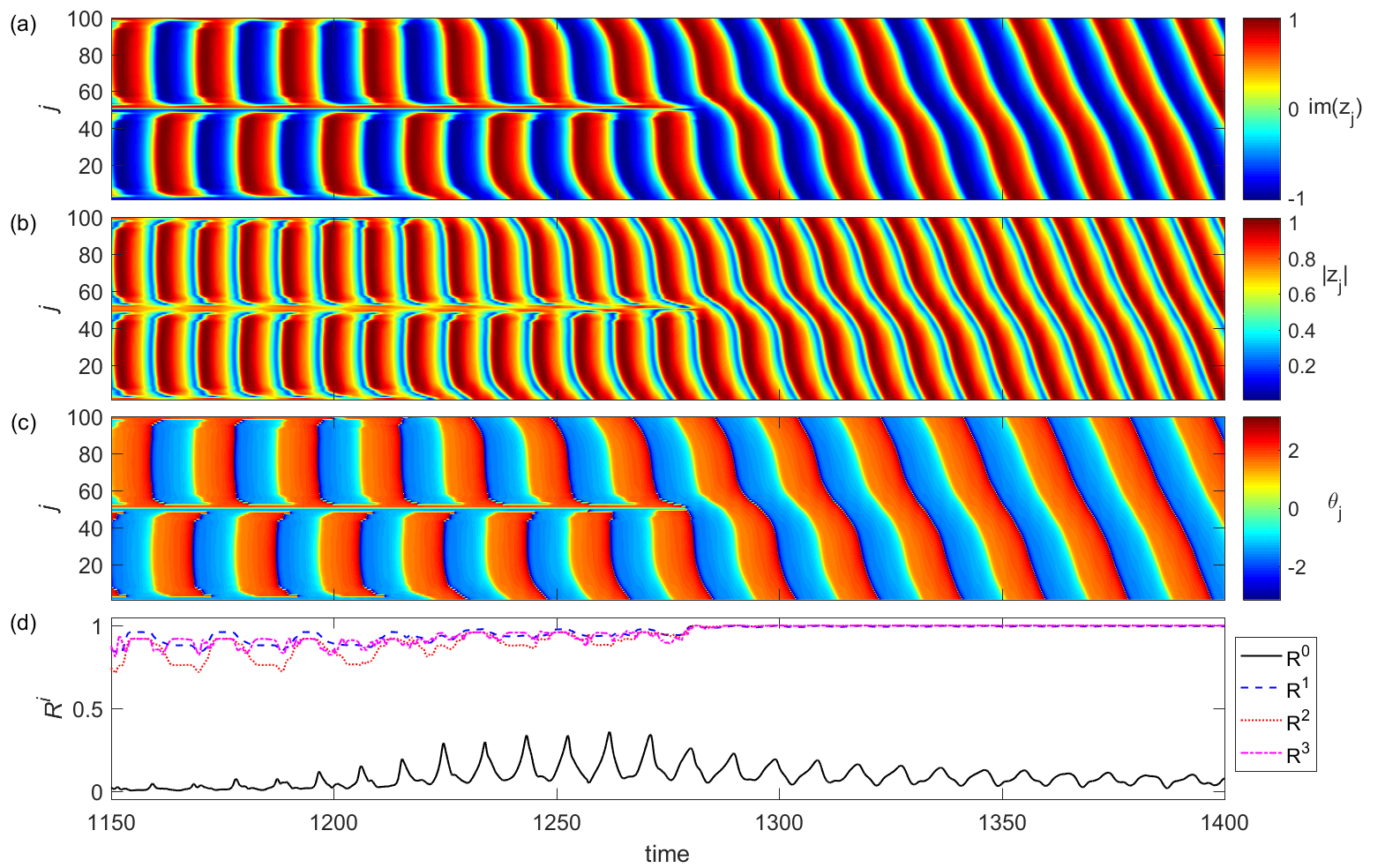}
\caption{(Color online) Space-time plots showing gradual collapse of amplitude chimera towards a phase-lag 
synchronized regime: 
(a) $y_j=\mathrm{Im}(z_j)$; (b) $|z_j|$; 
(c) $\theta_j=\arg(z_j)$; 
(d) Time-series of mean field parameters $R^0$ (solid, black), $R^1$ (dashed, blue), $R^2$ (dotted, red), $R^3$ (dash-dotted, magenta). 
Parameters: $P=5$, $\sigma=8$, $\tau=\pi$, $\lambda= 1$, $\omega = 2$, $N=100$.}
\label{fig2}
\end{figure*}

As an example we apply the mean field parameters to detect the transition from an amplitude chimera state to coherent dynamics. 
In Fig. \ref{fig2} we show the space-time plots of $\mathrm{Im}(z_j)$, $|z_j|$, $\mathrm{arg}(z_j)$, 
and the time evolution of $R^0$, $R^1$, $R^2$, and $R^3$  for the Stuart-Landau ring network 
of $N=100$ units with constant time-delayed coupling. 
The initial conditions are chosen as an amplitude chimera pattern as in Figs. \ref{fig4_5}a,b. This choice leads to a transient 
amplitude chimera state, which collapses towards a coherent traveling wave at $t\approx 1280$ (Fig. \ref{fig2})  that develops into a constant phase-lag synchronization pattern at $t>1500$. 
As discussed above, the zeroth-order parameter $R^0$ cannot be used to indicate the collapse of the chimera, since it shows irregular oscillatory behavior in time. In contrast, the higher order mean field parameters $R^1,R^2,\dots$, are highly sensitive to 
transitions from partial incoherence to various coherent wave-like structures, and provide an appropriate measure of the chimera collapse.

\section{The impact of time delay}
 
In this section, we analyze the lifetime of amplitude chimeras and the impact of time-delayed coupling on the network dynamics.
To determine the transient time, we use the first-order mean field parameter $R^1$ and employ the criterion $R^1 > 0.98$ to detect 
the transition to a coherent structure.

\subsection{Instantaneous coupling ($\tau=0$)}

First, for reference we show the map of regimes in Eq.(\ref{eq_sys}) for instantaneous coupling ($\tau=0$) in the plane of coupling range $P$ and coupling strength $\sigma$ in Fig. \ref{fig4}a. The two main regions are coherent patterns (SYNC) represented by in-phase and phase-lag synchronized oscillations (traveling waves), and chimera death states (CD) with different number of clusters. In the simulations, we take a ring of $N=100$ Stuart-Landau oscillators with $\lambda= 1$, $\omega = 2$, and integrate the system until $t=5000$ time units. As initial condition for the simulation we choose a snapshot of an amplitude chimera state for $P=1$ (see Figs. \ref{fig4_5}a,b). 
 
Fig.~\ref{fig5} depicts the lifetimes of amplitude chimeras indicated by the color code. Panels (a)-(f) correspond to the respective delay times of of Fig. \ref{fig4}. The black region marks chimera death states. In panel (a) ($\tau=0$) one can see that the lifetime of amplitude chimeras is significantly larger in the parameter region of low values of $P$ and large values of the coupling parameter $\sigma$ than everywhere else. Nevertheless, the lifetime of amplitude chimeras does not exceed $800$ time units ($\approx 255 T$) for the case of instantaneous coupling. This explains why the amplitude chimera regime is not visible in Fig. \ref{fig4}a after a simulation time of $t=5000$. 

Therefore, from the diagrams in Fig. \ref{fig4}a and Fig. \ref{fig5}a we conclude that without time delay in the coupling, 
the initial amplitude chimera state has a relatively short lifetime (at most a few hundred oscillation periods) and collapses into one of two different possible asymptotic states: asymptotic coherent states, in particular, in-phase synchronized oscillations (Figs. \ref{fig4_5}c,d), or chimera death states with several clusters (Figs. \ref{fig4_5}e,f).

\subsection{Constant time-delay coupling}

\begin{figure*}
\centering
\includegraphics[width=0.32\textwidth]{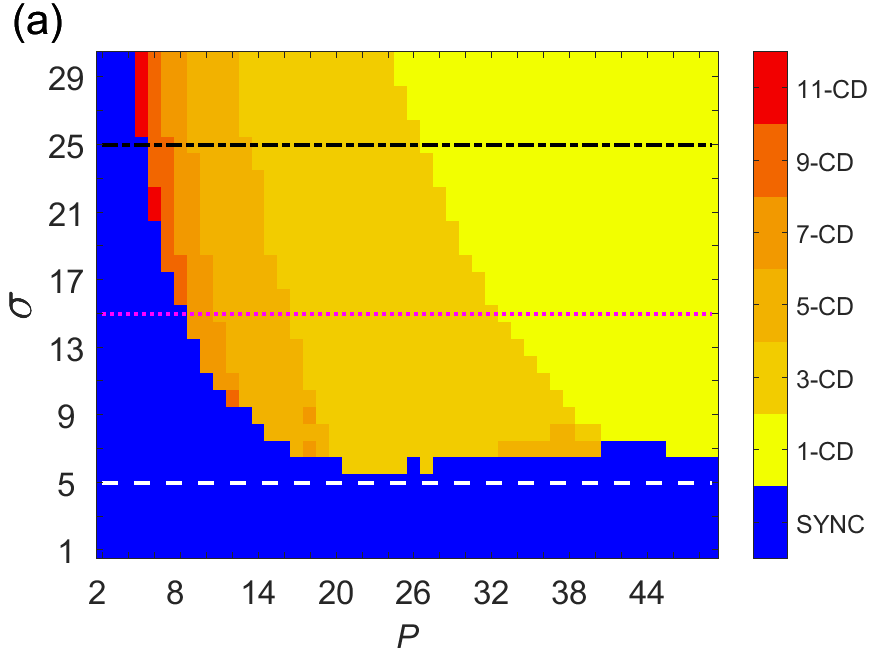}
\includegraphics[width=0.32\textwidth]{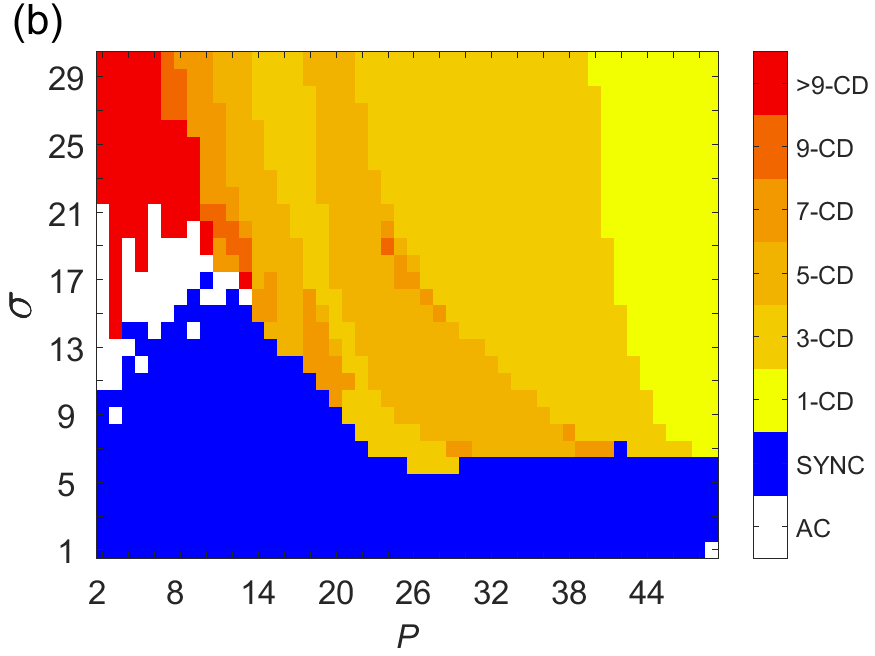}
\includegraphics[width=0.32\textwidth]{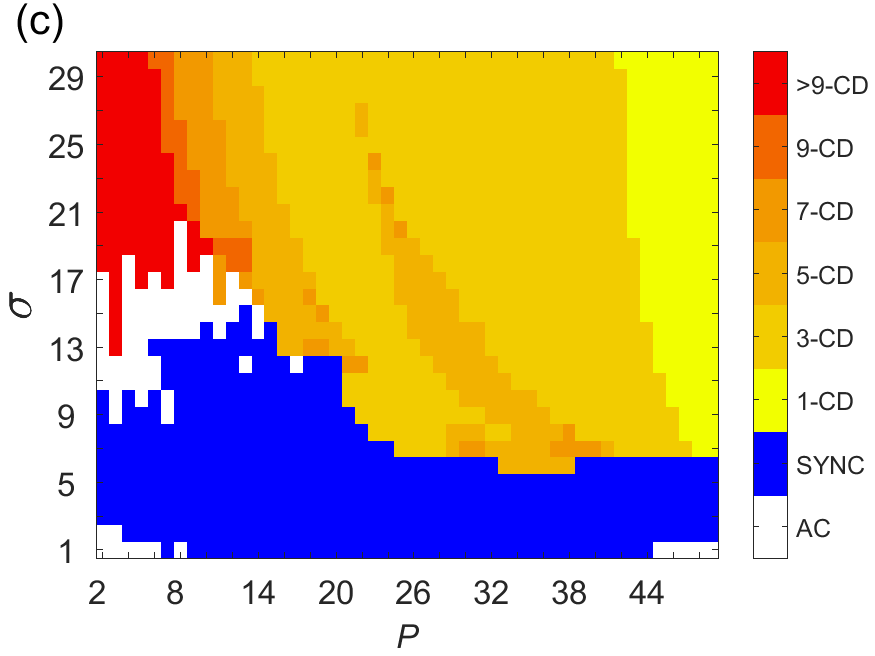}\\
\includegraphics[width=0.32\textwidth]{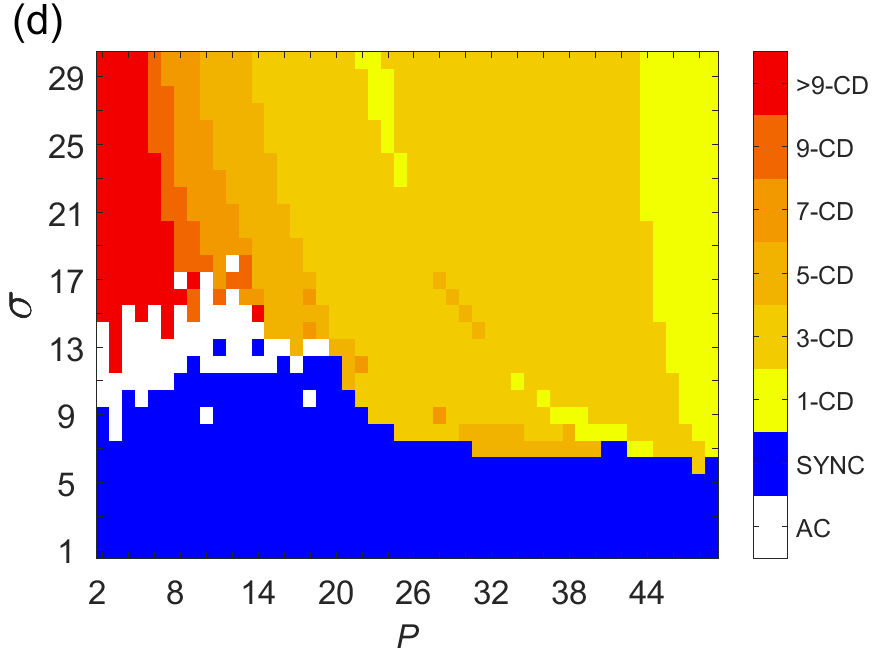}
\includegraphics[width=0.32\textwidth]{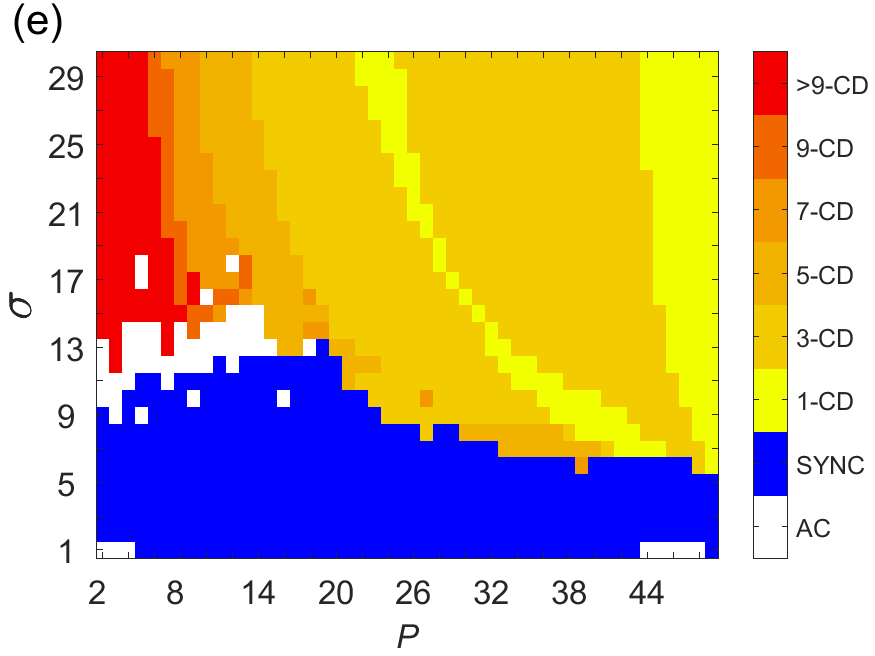}
\includegraphics[width=0.32\textwidth]{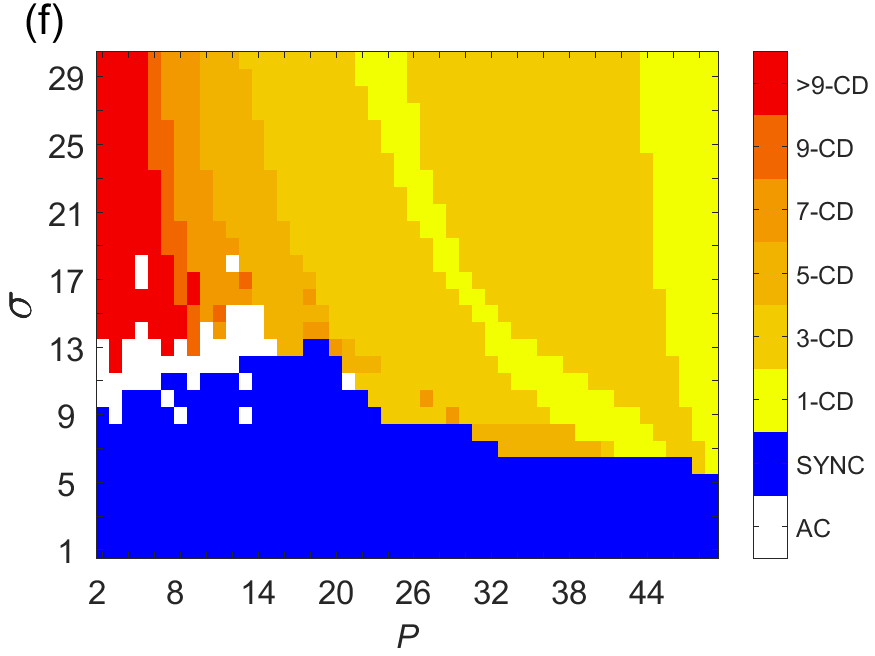}\\
\caption{(Color online) Map of dynamic regimes in a ring-network of $N=100$ delay-coupled Stuart-Landau oscillators in the plane of coupling range $P$ and coupling strength $\sigma$ for constant values of the time delay: 
(a) $\tau = 0$;  (b) $\tau=\pi/4$; (c) $\tau=\pi/2$; (d) $\tau=\pi$; (e) $\tau=3\pi/2$; (f) $\tau=2\pi$.
Color code: 1-CD: 1-cluster chimera death; 3-CD: 3-cluster chimera death; n-CD: n-cluster chimera death; 
SYNC: coherent states (synchronized oscillations, traveling waves, etc.); AC: amplitude chimera and related incoherent states. 
Other parameters: $\lambda= 1$, $\omega = 2$. Simulation time $t=5000$.}
\label{fig4}
\end{figure*}

\begin{figure*}
\centering
\includegraphics[width=0.32\textwidth]{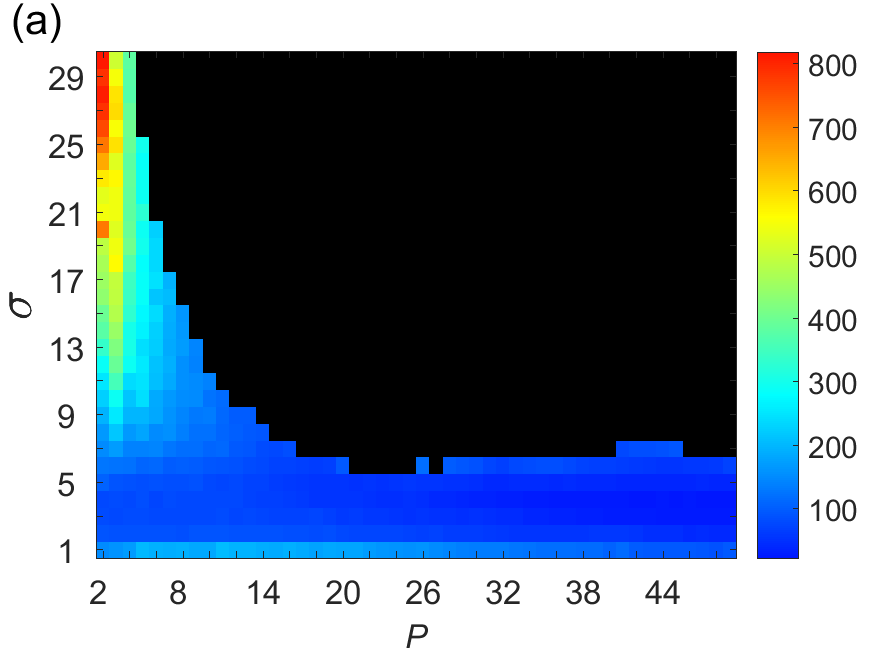}
\includegraphics[width=0.32\textwidth]{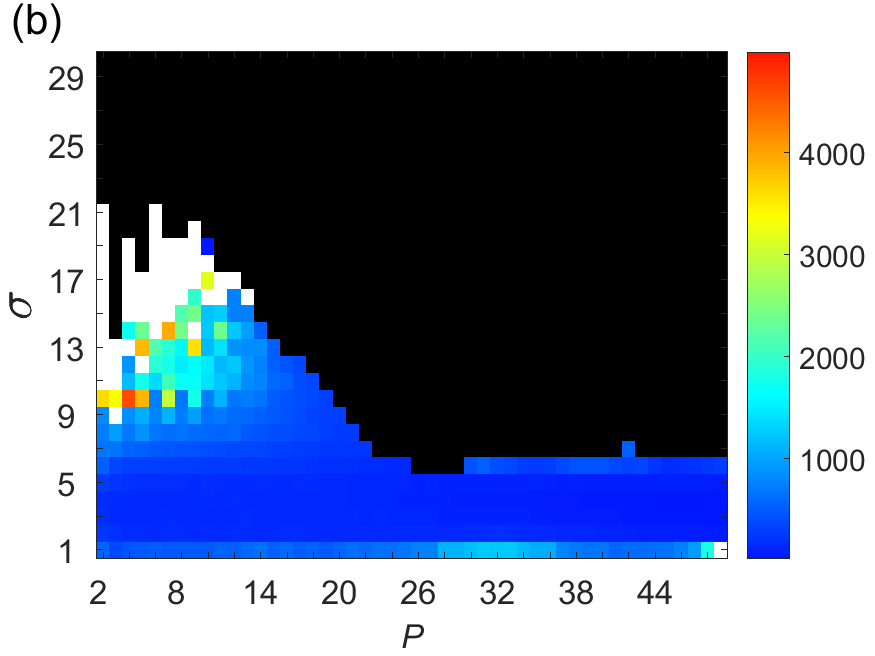}
\includegraphics[width=0.32\textwidth]{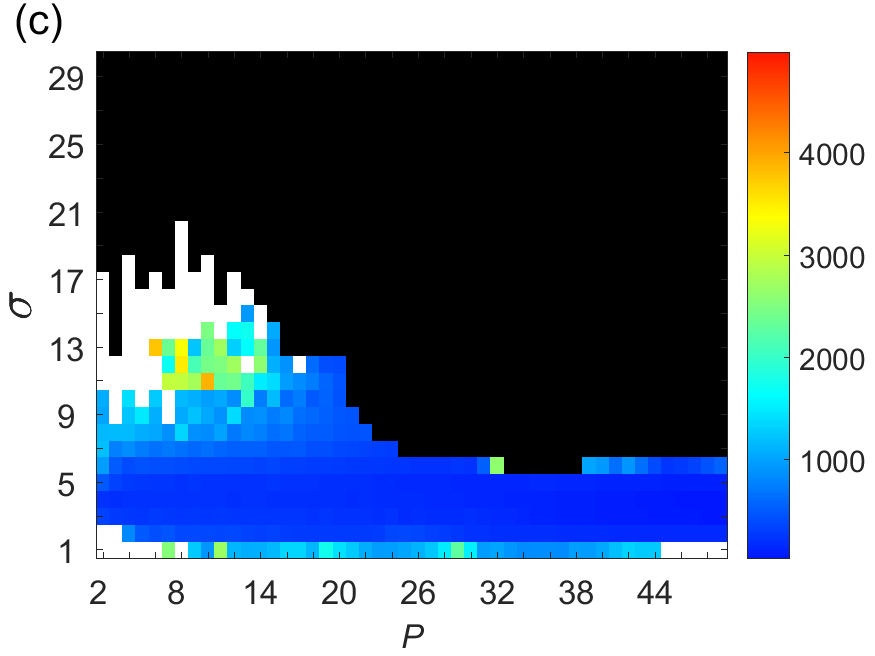}\\
\includegraphics[width=0.32\textwidth]{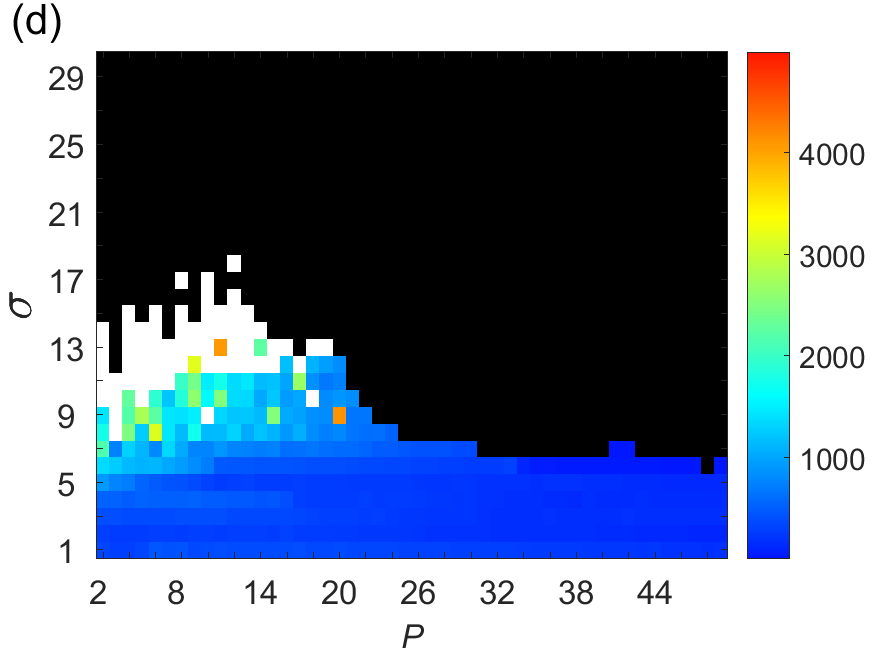}
\includegraphics[width=0.32\textwidth]{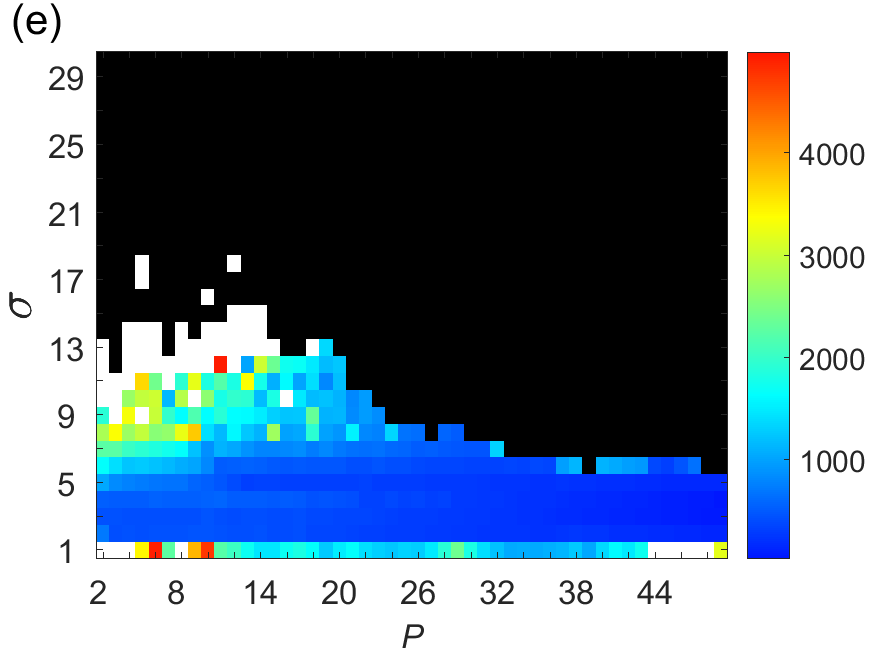}
\includegraphics[width=0.32\textwidth]{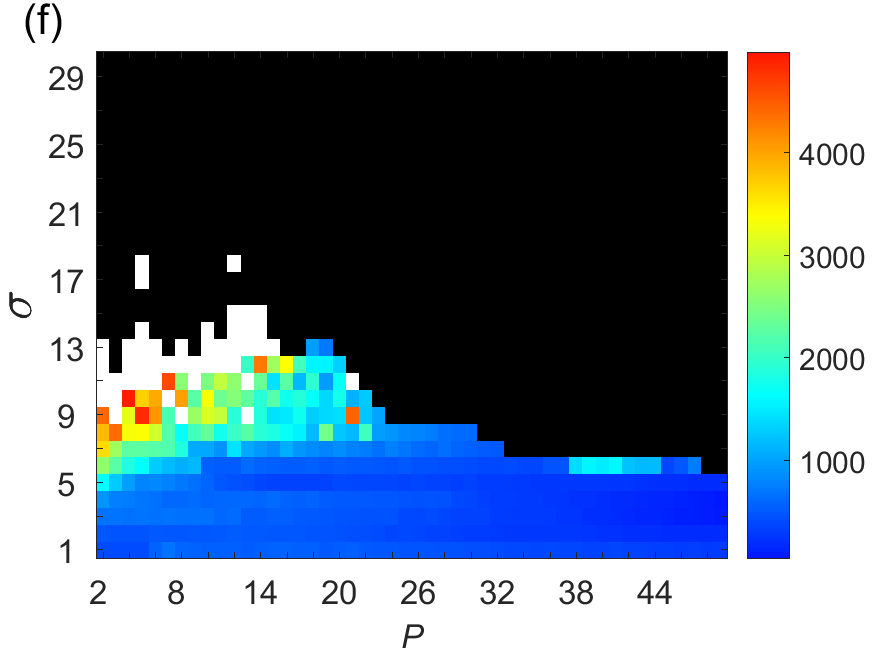}\\
\caption{(Color online) Lifetime of amplitude chimeras in the plane of coupling range $P$ and coupling strength 
$\sigma$, corresponding to Fig. \ref{fig4}.
Color code indicates the time of transition from partially incoherent states (amplitude chimera) to coherent states; 
the white region denotes amplitude chimeras and related incoherent states living longer than $t=5000$, and the black region shows 
stable steady states (chimera death states). Other parameters as in Fig. \ref{fig4}.}
\label{fig5}
\end{figure*}

\begin{figure}
\centering
$\begin{array}{cc}
\includegraphics[width=0.49\columnwidth]{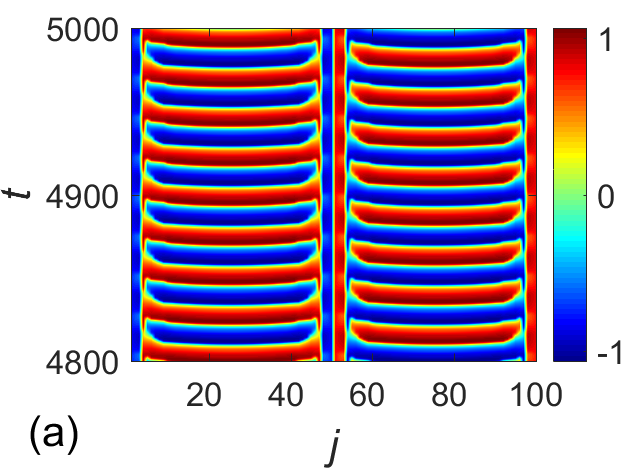}&
\includegraphics[width=0.49\columnwidth]{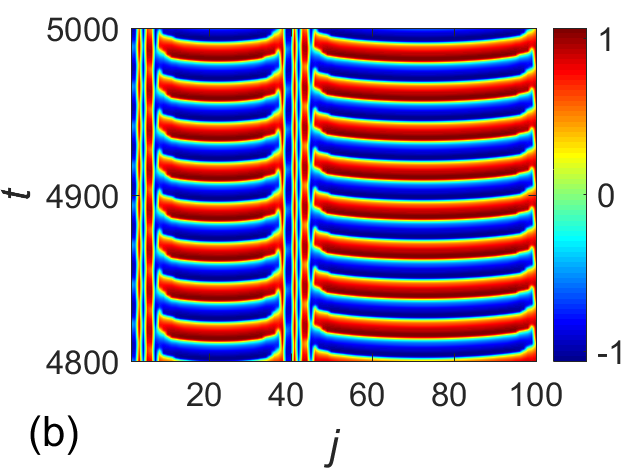}\\
\includegraphics[width=0.49\columnwidth]{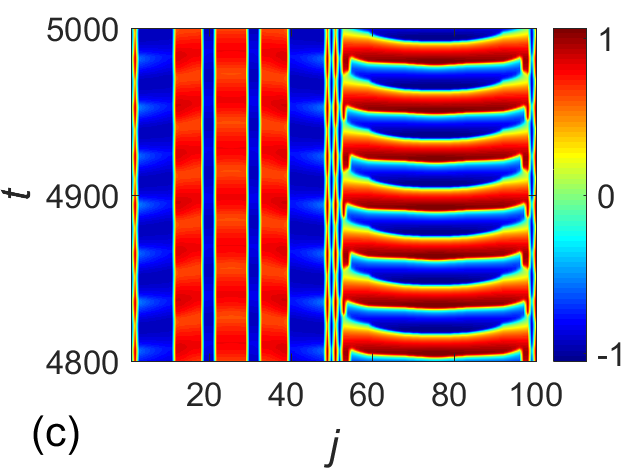}&
\includegraphics[width=0.49\columnwidth]{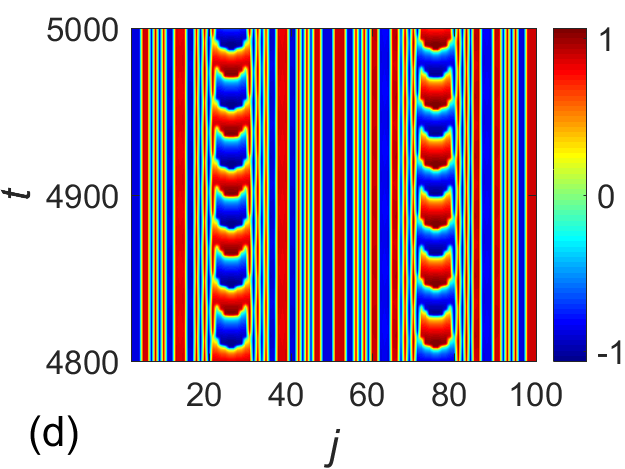}
\end{array}$
\caption{(Color online) Examples of partially incoherent space-time patterns related to amplitude chimera states for constant time delay $\tau=\pi$ in the coupling and integration time $t=5000$:
(a) Symmetric amplitude chimera, $\sigma=11$, $P=5$; 
(b) Asymmetric amplitude chimera, $\sigma=11$, $P=4$;
(c) Partial amplitude chimera, $\sigma=13$, $P=19$;
(d) Partial oscillation death, $\sigma=15$, $P=4$.
Other parameters: $\lambda= 1$, $\omega = 2$, $N = 100$.}
\label{fig5_6}
\end{figure}

Including time delay in the coupling changes both the dynamical regimes of the network and the lifetime of 
amplitude chimeras and related partially incoherent patterns. We investigate the states of the network Eq. (\ref{eq_sys}) for increasing delay values: $\tau=\pi/4$; $\pi/2$; $\pi$; $3\pi/2$; $2\pi$ (Fig. \ref{fig4}b--f, respectively), where the delay times are chosen as integer or non-integer multiples of the intrinsic period of the system $T=\pi$. The corresponding lifetime diagrams are shown in Fig. \ref{fig5}b--f, respectively, where the color code indicates the lifetime of the amplitude chimera (AC) patterns. Initial conditions in our simulations are snapshots of amplitude chimeras and the history function for $t<0$ is taken to be constant and equal to the initial condition at $t=0$. The presence of delay in the coupling significantly prolongs the lifetime of amplitude chimeras: they continue to exist at $t = 5000$ for certain parameter values $\sigma$ and $P$ (white regions in panels (b)--(f) of Figs. \ref{fig4} and \ref{fig5}). 
Time delay induces various partially incoherent states related to amplitude chimeras which are characterized by relatively long lifetimes (Fig. \ref{fig5_6}). These spatio-temporal patterns include a typical symmetric amplitude chimera with two coherent domains of the same spatial width oscillating in antiphase, separated by spatially incoherent regions consisting typically of few oscillators (Fig. \ref{fig5_6}a). For other parameters, the amplitude chimera pattern can be asymmetric, consisting of two phase-antiphase coherent domains with different spatial width (Fig. \ref{fig5_6}b), or one of the coherent domains can collapse into an oscillation death state (Fig. \ref{fig5_6}c), or the width of coherent oscillating domains can be significantly smaller than the oscillation death region (Fig. \ref{fig5_6}d). Similar partially incoherent patterns have been found for other models \cite{BAN15,DUT15,BAN16}.

The amplitude chimera region in Figs. \ref{fig4} and \ref{fig5} appears already for small delay values, replacing the region of synchronized states for small $P$ and large $\sigma$, and shifting towards the intermediate range of $\sigma$ values as the delay is increased. Gradually, it evolves into an irregular region positioned around the middle of the $\sigma$ interval and the first third of the $P$ interval ($P<20$). At the same time, for increasing $\tau$ the in-phase synchronized state observed for small $P$ and large $\sigma$ is replaced by chimera death patterns with large number of clusters (red region in Fig. \ref{fig4}). 
Amplitude chimera states still exist as the delay time becomes larger, without significantly changing their 
position further and exhibiting an occasional subsidiary appearance at low values of $\sigma$ for different number of 
nearest neighbors within the interval $P\in[2,20]$. 

\begin{figure}
\centering
\includegraphics[width=0.9\columnwidth]{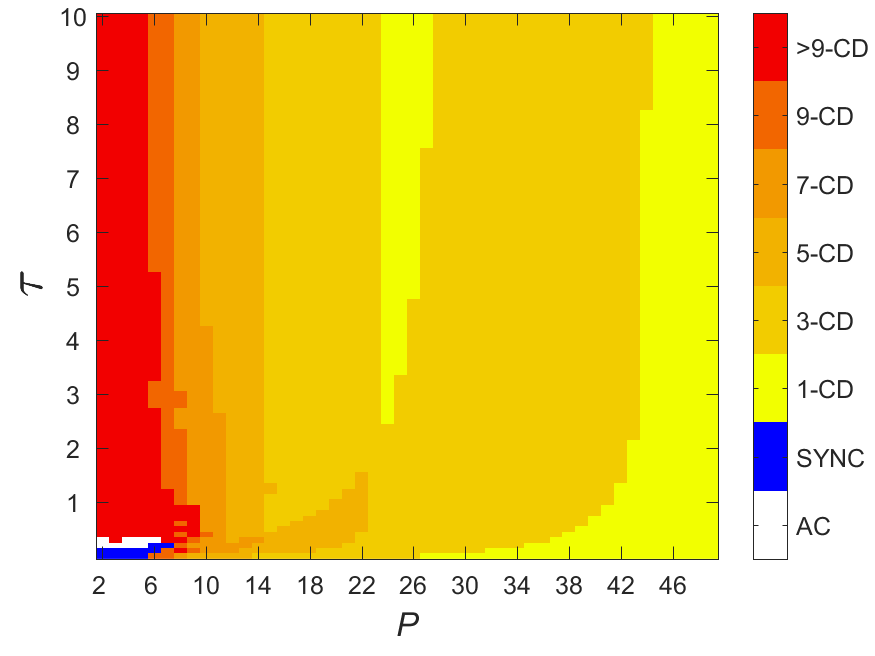}
\caption{(Color online) Map of dynamic regimes in the plane of coupling range $P$ and time delay $\tau$ 
for coupling strength $\sigma=25$. 
Color code: 1-CD: 1-cluster chimera death; 3-CD: 3-cluster chimera death; n-CD: n-cluster chimera death. 
SYNC: coherent states (synchronized oscillations, traveling waves, etc.). 
AC: amplitude chimeras and related partially incoherent states. Other parameters: $\lambda= 1$, $\omega = 2$, $N = 100$.}
\label{fig6}
\end{figure}

This trend of changing the network states by delay can be better understood by analyzing the dynamical 
states in the ($P,\tau$)-plane for different values of the coupling strength $\sigma$. 
For that purpose, we fix $\sigma$ at three representative values, indicated by the horizontal lines in Fig. \ref{fig4}a 
(dashed, white line for $\sigma=5$; dotted, red line for $\sigma=15$; dash-dotted, black line for $\sigma=25$).
In Fig. \ref{fig6} we depict  the ($P,\tau$) diagram for $\sigma = 25$. It clearly indicates that the region of synchronized states for low $P$ values is rapidly replaced by amplitude chimeras for increasing time delay. Furthermore, amplitude chimeras disappear at $\tau\approx 0.4$, and the region of low $P$ values is taken over by chimera death states with the number of clusters in the coherent domain exceeding 9. Simultaneously, the region of 1-cluster chimera death (1-CD) is strongly reduced and partially replaced by 3-cluster 
chimera death.   

\begin{figure}
\centering
\includegraphics[width=0.9\columnwidth]{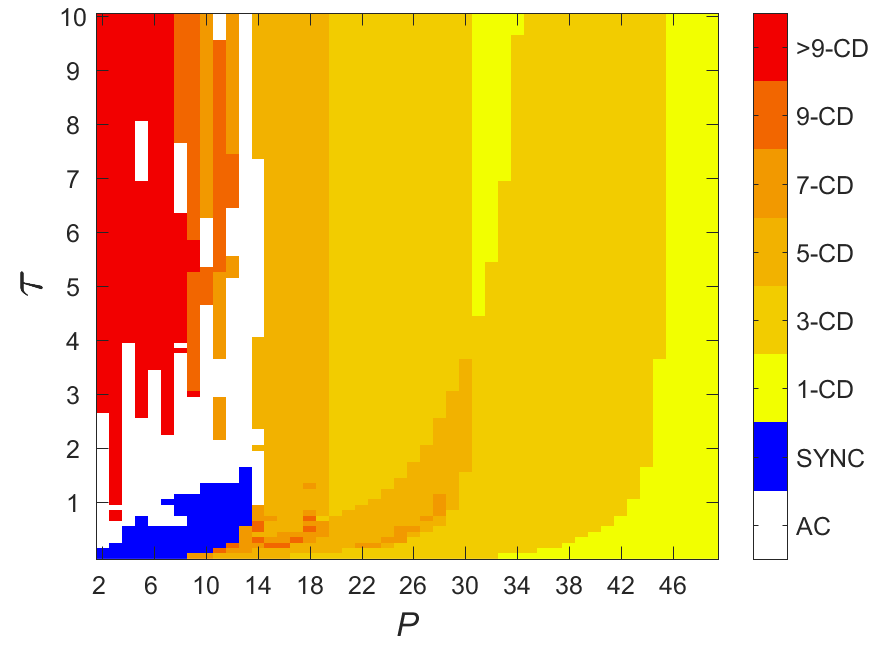}\\
\caption{(Color online) Same as Fig.~\ref{fig6} for coupling strength $\sigma=15$.}
\label{fig7}
\end{figure}

When the coupling between the elements of the network is weaker ($\sigma=15$), the synchronization region for $P<14$ starts to transform into the amplitude chimera regime already for $\tau\approx 0.2$ (Fig. \ref{fig7}). For increasing $\tau$, amplitude chimeras still exist in this region, being further partially replaced by chimera death states with large number of clusters.

\begin{figure}
\centering
\includegraphics[width=0.9\columnwidth]{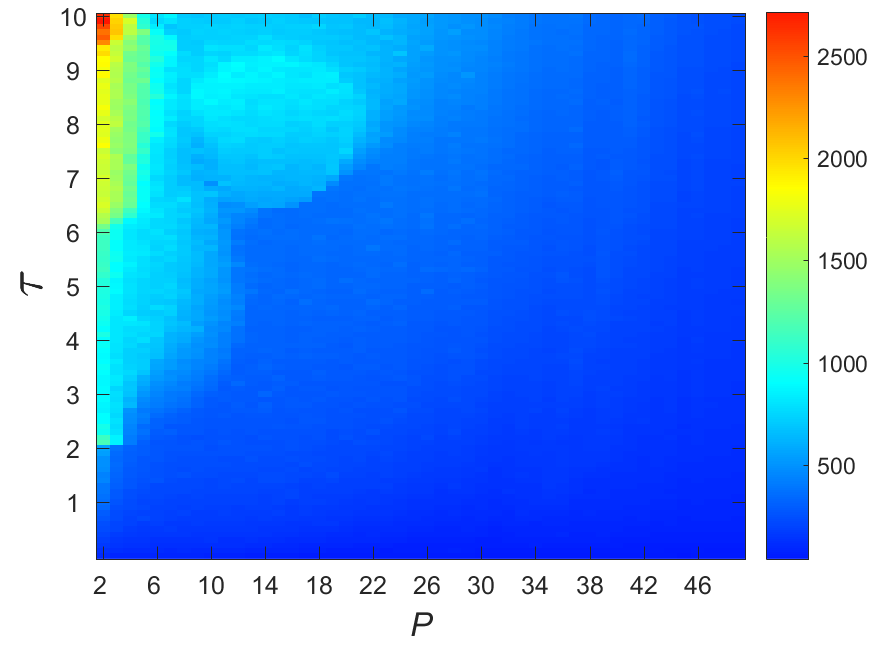}\\
\caption{(Color online) Lifetime of amplitude chimeras in the plane of coupling range $P$ and time delay $\tau$ for coupling strength $\sigma=5$. Color code indicates the time of transition from partially incoherent states (amplitude chimera) 
to coherent states (in-phase synchronization). Other parameters: $\lambda= 1$, $\omega = 2$, $N = 100$.}
\label{fig8}
\end{figure}

To investigate the influence of time-delayed coupling on the synchronized dynamics, we calculate 
the lifetime of amplitude chimera states for $\sigma=5$ in the ($P,\tau$)-plane (Fig. \ref{fig8}). The transient times towards the synchronized regime become larger with increasing delay for each $P$, although this enlargement is more rapid for lower values of $P$. 

\begin{figure*}
\centering
\includegraphics[width=0.32\textwidth]{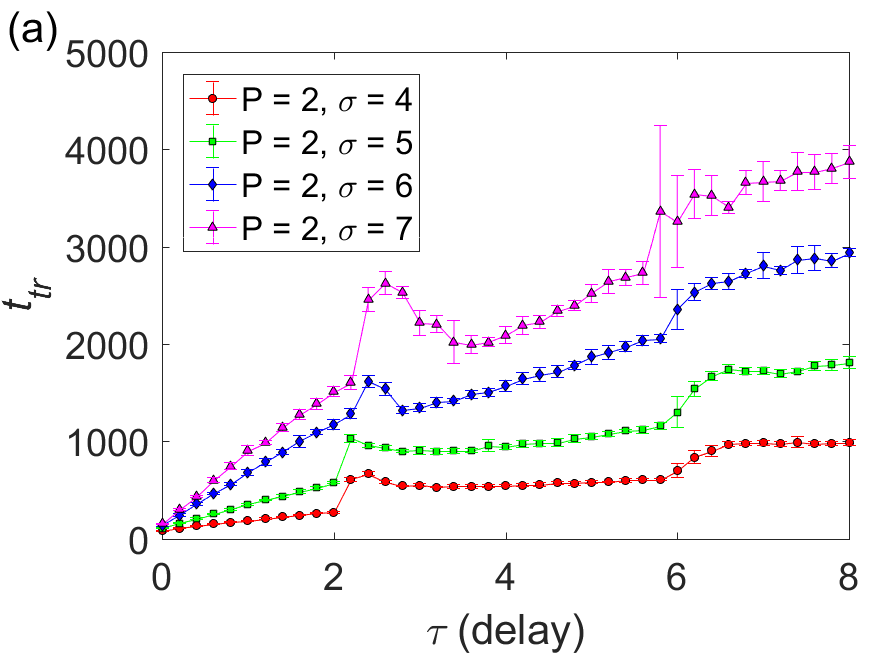}
\includegraphics[width=0.32\textwidth]{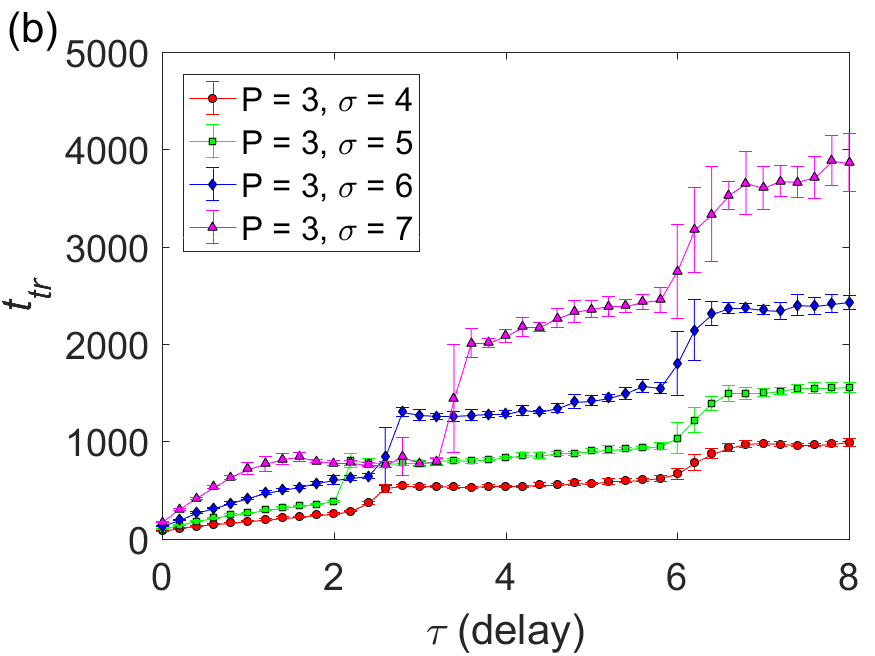}
\includegraphics[width=0.32\textwidth]{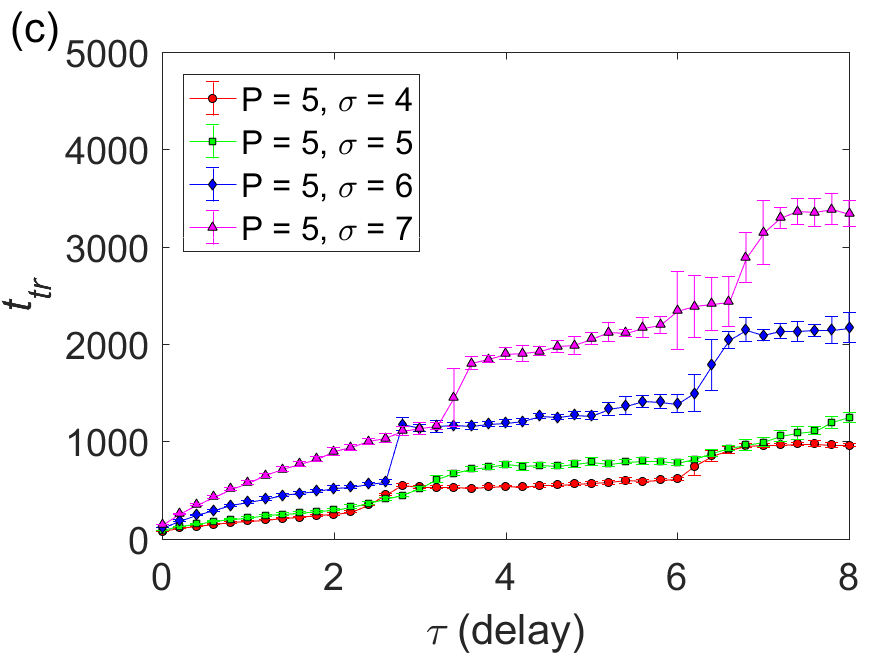}
\caption{(Color online) Lifetime of amplitude chimeras $t_{tr}$ in dependence on time delay $\tau$ for four selected values 
of coupling strength ($\sigma = 4, 5, 6, 7$) averaged over ten different initial conditions that favor amplitude
chimeras. The number of nearest neighbors: (a) $P=2$; (b) $P=3$; (c) $P=5$. 
Other parameters: $\lambda= 1$, $\omega = 2$, $N=100$.}
\label{fig9}
\end{figure*}

The same trend of the lifetime prolongation for the amplitude chimera state by time delay is generally found 
for other values of coupling strength $\sigma$. In panels (a)--(c) of Fig. \ref{fig9} we calculate the lifetime 
of amplitude chimeras in dependence on time delay for different values of coupling strength $\sigma = 4, 5, 6, 7$. We compare the results obtained for three values of the coupling range: $P=2$, $P=3$, $P=5$ (Fig. \ref{fig9}a,b,c, respectively). The lifetime of amplitude chimeras generally grows faster when the number of nearest neighbors in the network $P$ is low (Fig. \ref{fig9}a). 
We conclude that by appropriately choosing the values of time delay one can extend the lifetime of amplitude 
chimeras up to a desired value.

\begin{figure*}
\centering
\includegraphics[width=\textwidth]{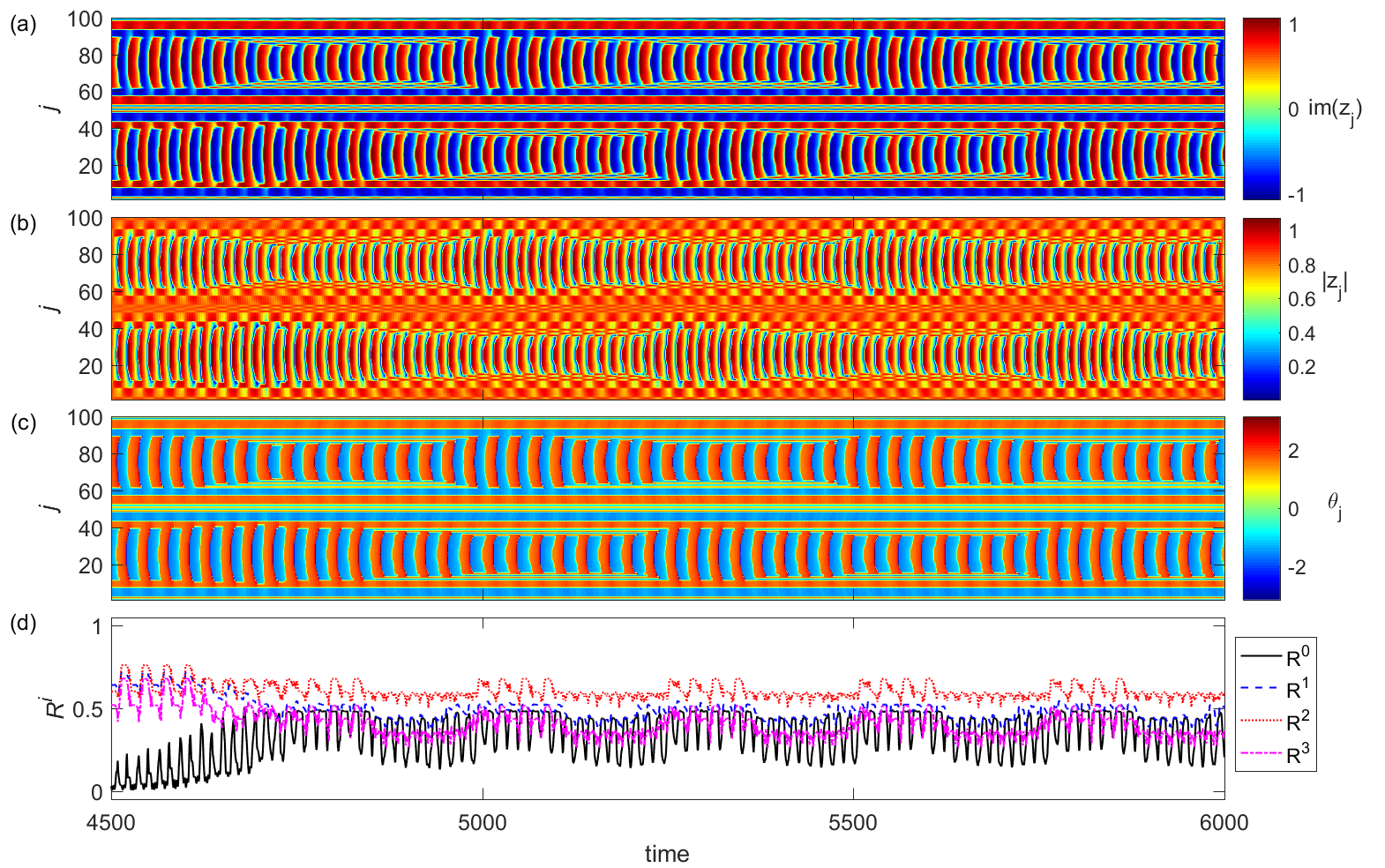}
\caption{(Color online) Space-time plots for a long-living ``breathing'' amplitude chimera.  
(a) $y_j=\mathrm{im}(z_j)$; (b) $|z_j|$; 
(c) $\theta_j=\arg(z_j)$; 
(d) Time-series of mean field parameters $R^0$ (solid, black), $R^1$ (dashed, blue), $R^2$ (dotted, red), $R^3$ (dash-dotted, magenta). 
Parameters: $P=12$, $\sigma=13$, $\tau=\pi$, $\lambda= 1$, $\omega = 2$, $N = 100$.}
\label{fig9_10}
\end{figure*}

In the parameter region where the network dynamics is represented by amplitude chimera-related states,
we find a peculiar delay-induced pattern of a ``breathing'' amplitude chimera, which starts at around $t\approx 4500$ and
lasts much longer than the simulation time indicated in Fig. \ref{fig9_10}. 
We investigate the space-time plot of $\mathrm{Im}(z_j)$, $|z_j|$, $\mathrm{arg}(z_j)$ and the temporal development of the global order parameters $R^0,\dots R^3$ for a ``breathing'' amplitude chimera pattern with two coherent and two incoherent domains.
The size (spatial width) of the two coherent domains of the amplitude chimera is changing in time in a periodic manner (Fig. \ref{fig9_10}a,b,c). Moreover, these oscillations occur in anti-phase for the two coherent domains, i.e., when one coherent cluster attains a maximum width, the other has a minimum, and vice versa. The periodicity is inherited also in the time evolution of the order parameters (Fig. \ref{fig9_10}d).  Specifically, we have integrated the system until $t=20000$ time units ($\approx 6370 T$) and observe a sustained ``breathing'' amplitude chimera, with a breathing period of each coherent domain approximately equal to $500$ time units ($\approx 159 T$).

The results we have discussed above are obtained for a special initial condition resulting in the amplitude chimera pattern with two equally sized spatially coherent regions (symmetric amplitude chimera). However, increasing the lifetime of amplitude chimeras by introducing time delay in the coupling is not confined to these special initial conditions. The effect of an essential enhancement of the chimera lifetime by time delay suggests a possibility to design a desired multicluster as well as asymmetric amplitude chimera states by appropriately choosing initial conditions and making these pattern long-lasting by adding delay in the coupling.

\begin{figure}
\centering
$\begin{array}{cc}
\includegraphics[width=0.49\columnwidth]{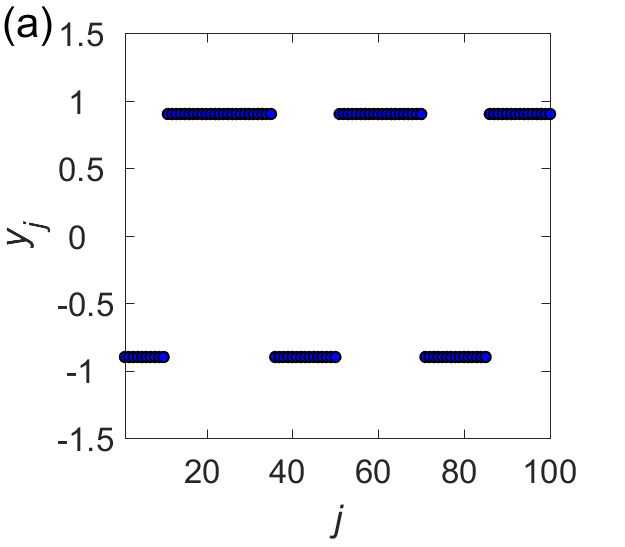}&
\includegraphics[width=0.41\columnwidth]{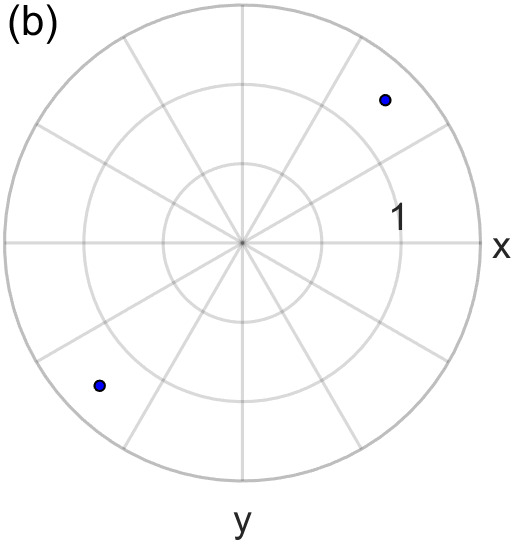}\\
\includegraphics[width=0.49\columnwidth]{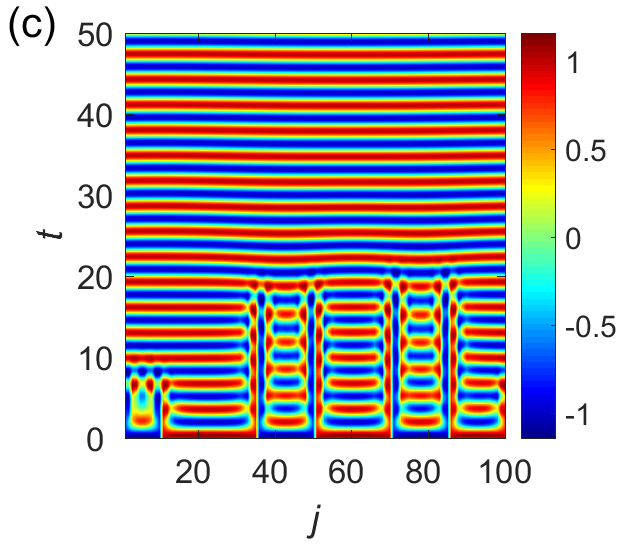}&
\includegraphics[width=0.49\columnwidth]{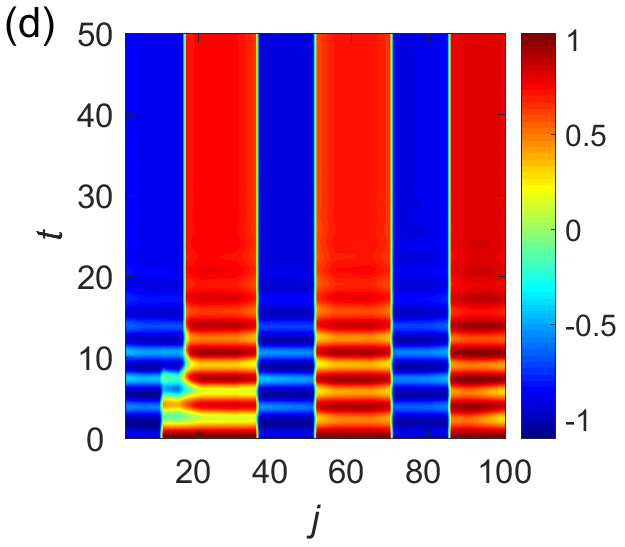}
\end{array}$
\caption{(Color online) Asymmetric initial condition ($t\leq 0$): (a) Snapshot of the variable $y_j=\mathrm{Im}(z_j)$, (b) phase portrait in the complex $z$-plane. (c) Space-time plot of $y_j=\mathrm{Im}(z_j)$ showing the collapse of initial asymmetric multicluster amplitude chimera towards in-phase synchronized regime at $t\approx 20$ for $P=2$ and $\sigma=14$. 
(d) Collapse into an asymmetric 3-cluster oscillation death state for $P=20$, $\sigma=12$. Other parameters: $\tau=0$, $\lambda= 1$, $\omega = 2$, $N = 100$.}
\label{fig9_10_2}
\end{figure}

\begin{figure*}
\centering
\includegraphics[width=\textwidth]{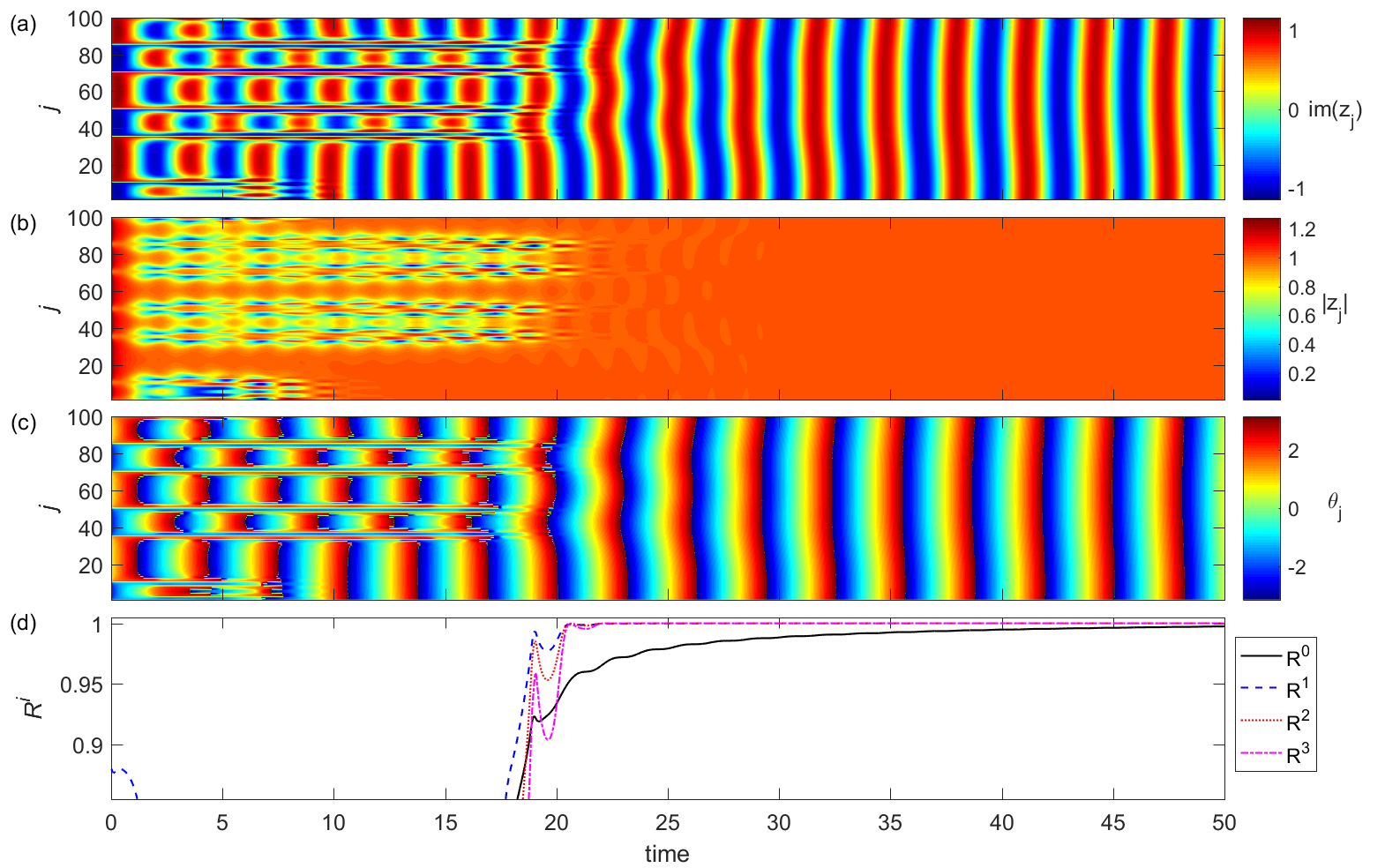}
\caption{(Color online) Space-time plots showing gradual collapse of an asymmetric multicluster amplitude chimera towards a zero-lag 
synchronized state: 
(a) $y_j=\mathrm{Im}(z_j)$; 
(b) $|z_j|$; 
(c) $\theta_j=\arg(z_j)$; 
(d) Time-series of mean field parameters $R^0$ (solid, black), $R^1$ (dashed, blue), $R^2$ (dotted, red), $R^3$ (dash-dotted, magenta). 
Parameters: $P=2$, $\sigma=14$, $\tau=0$, $\lambda= 1$, $\omega = 2$, $N = 100$.}
\label{fig_9_10_3}
\end{figure*}

As an example, we choose spatially asymmetric initial conditions (Figs. \ref{fig9_10_2}a and \ref{fig9_10_2}b). In the absence of coupling delay, the asymmetric multicluster amplitude chimera which evolves from these initial conditions dies out very fast, typically lasting only for a few time units (Fig. \ref{fig9_10_2}c). Depending on the parameters $P$ and $\sigma$, the final state is either in-phase synchronized (Fig. \ref{fig9_10_2}c), or multicluster oscillation death (Fig. \ref{fig9_10_2}d). 
The transition between the amplitude chimera and zero-lag synchronization for instantaneous coupling ($\tau=0$), $P=2$ and $\sigma=14$ occurs gradually, starting from $t\approx 20$, 
where the amplitude chimera collapses, forming a more complicated coherent waveform that subsequently develops into a complete zero-lag synchronization pattern around $t\approx 50$ (Fig. \ref{fig_9_10_3}).

\begin{figure*}
\centering
\includegraphics[width=0.32\textwidth]{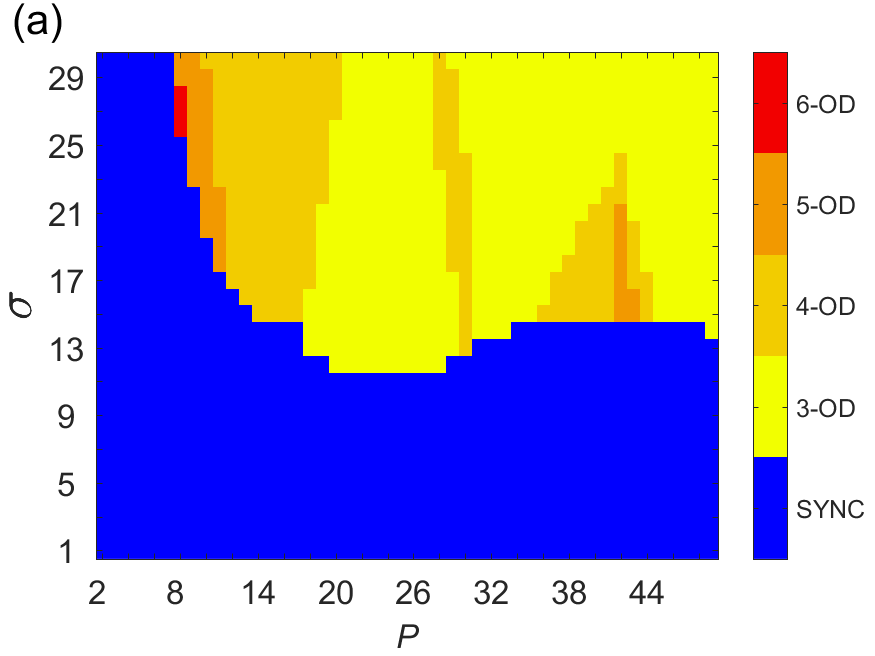}
\includegraphics[width=0.32\textwidth]{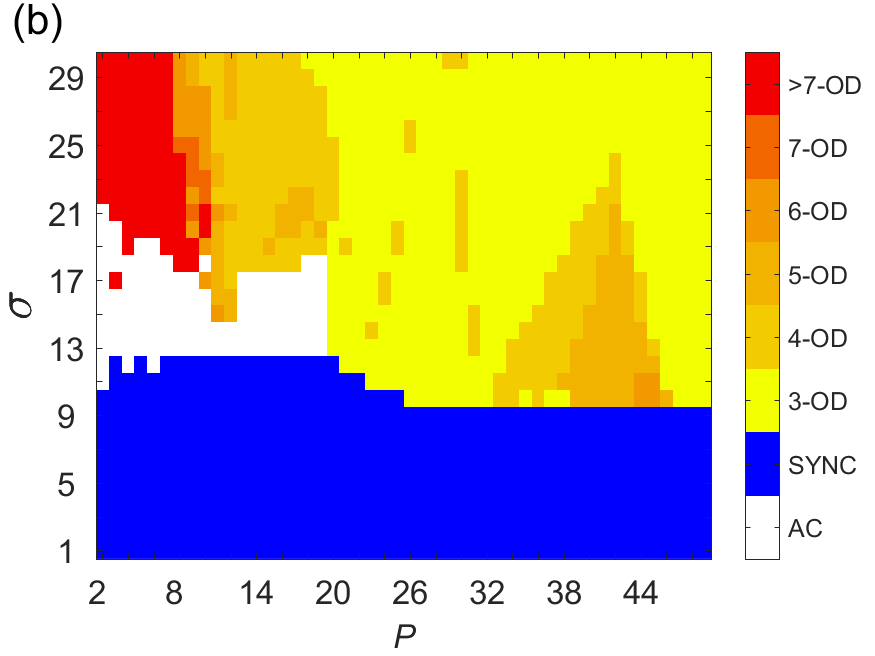}
\includegraphics[width=0.32\textwidth]{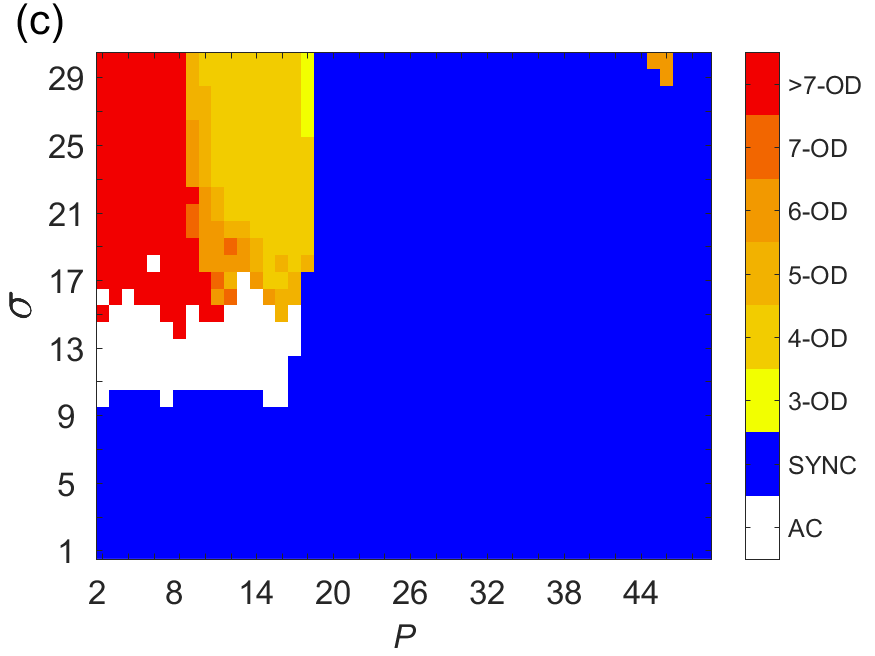}\\
\includegraphics[width=0.32\textwidth]{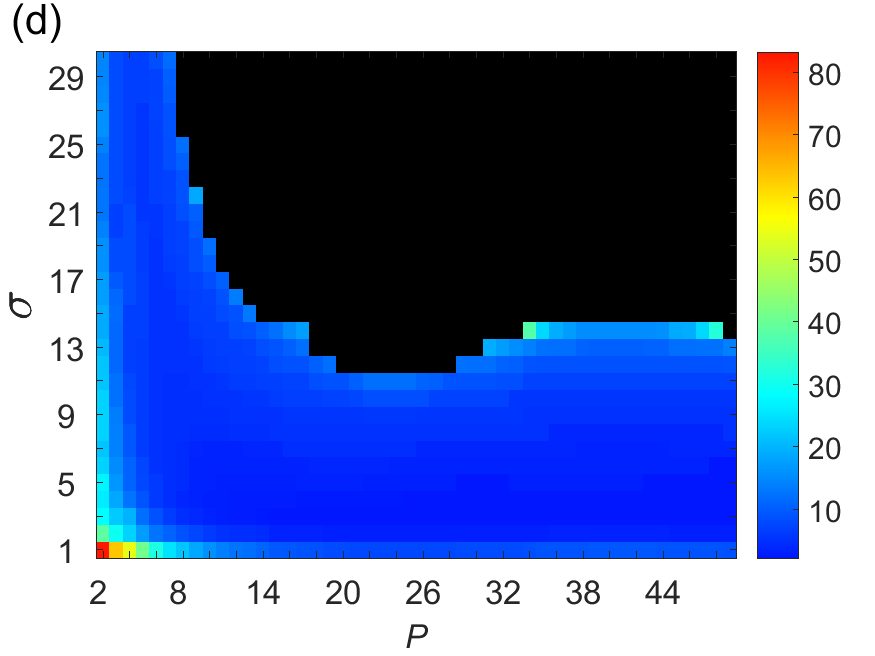}
\includegraphics[width=0.32\textwidth]{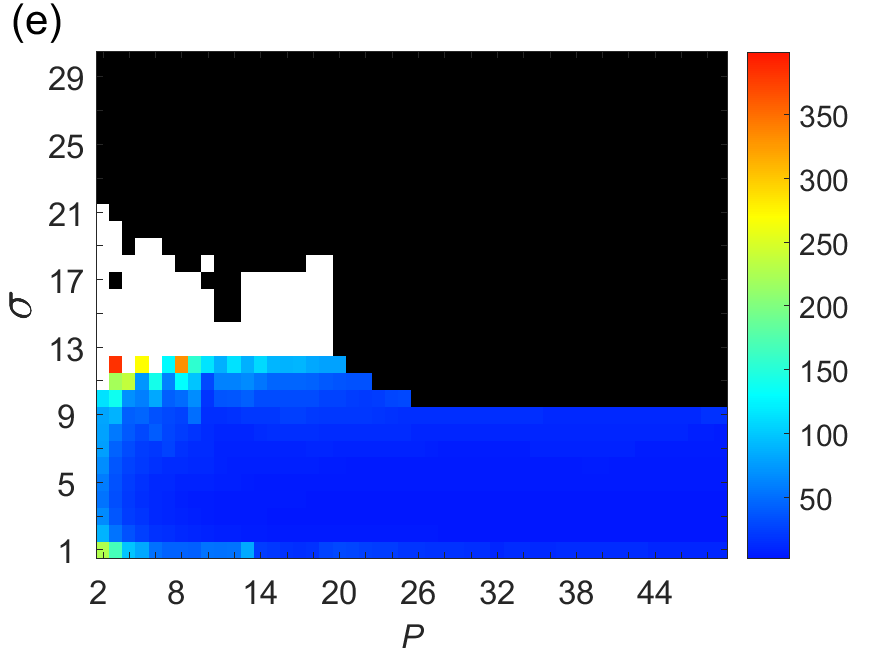}
\includegraphics[width=0.32\textwidth]{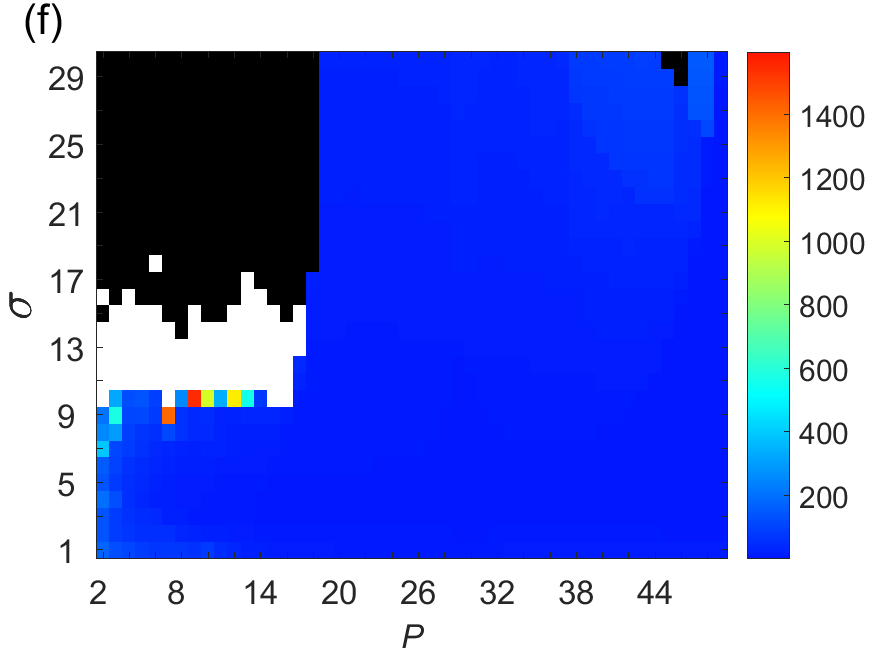}
\caption{(Color online) Map of dynamic regimes (a)--(c) and corresponding chimera lifetimes (d)--(f) for 
multicluster amplitude chimeras in the plane of coupling range $P$ and coupling strength $\sigma$ for constant 
delay coupling. Time delay: (a), (d) $\tau = 0$;  (b), (e) $\tau=\pi/4$; (c), (f) $\tau=\pi$.
Color scale in (a)--(c): 3-OD: 3-cluster oscillation death; 4-OD: 4-cluster oscillation death; n-OD: n-cluster oscillation death. 
SYNC: coherent states (in-phase synchronized oscillations, traveling waves, etc.). 
AC: multicluster amplitude chimera and related partially incoherent states. Color code in lifetime diagrams (d)--(f) 
indicates the time of transition from partially incoherent states (multicluster amplitude chimera) to coherent states. 
The white region denotes amplitude chimeras and related partially incoherent states; the black region denotes 
stable steady states (oscillation death). Other parameters: $\lambda= 1$, $\omega = 2$, $N = 100$.  Simulation time $t=5000$.}
\label{fig9_10_4}
\end{figure*}

\begin{figure}
\centering
$\begin{array}{cc}
\includegraphics[width=0.49\columnwidth]{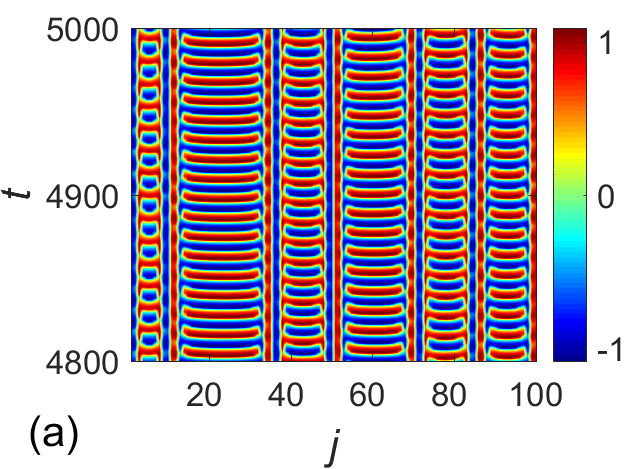}&
\includegraphics[width=0.49\columnwidth]{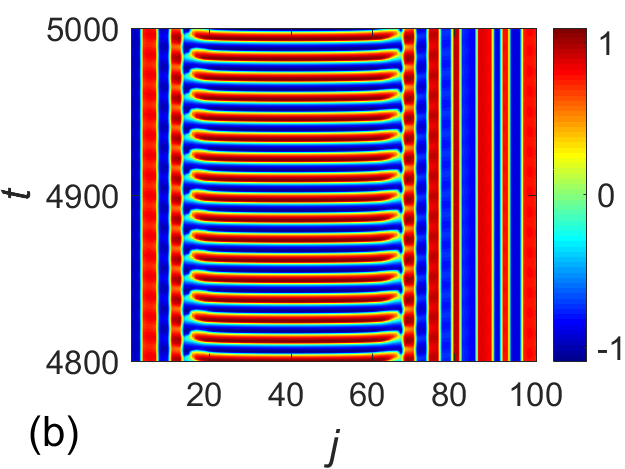}\\
\includegraphics[width=0.49\columnwidth]{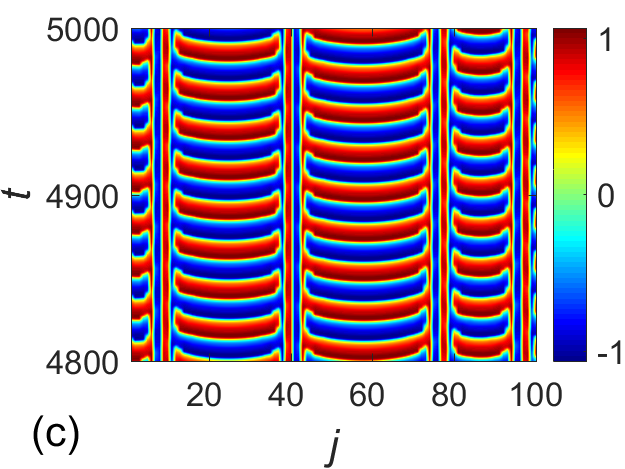}&
\includegraphics[width=0.49\columnwidth]{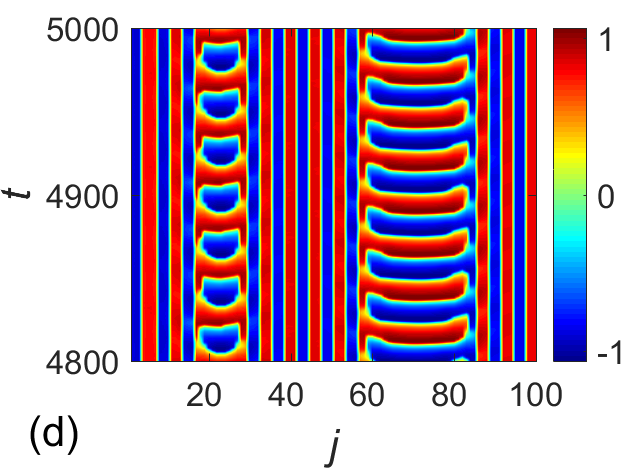}
\end{array}$
\caption{(Color online) Examples of partially incoherent dynamical patterns related to multicluster amplitude chimeras that survive the simulation time for constant time delay $\tau$ in the coupling. Parameters:
(a) $\sigma=13$, $P=2$, $\tau=\pi/4$; 
(b) $\sigma=14$, $P=5$, $\tau=\pi/4$;
(c) $\sigma=11$, $P=2$, $\tau=\pi$;
(d) $\sigma=12$, $P=5$, $\tau=\pi$.
Other parameters as in Fig. \ref{fig9_10_4}.}
\label{multicluster_examples}
\end{figure}

To gain insight into the influence of time delay on the 
lifetime of multicluster amplitude chimera states we analyze the map of regimes (Fig. \ref{fig9_10_4}a--c) and
the corresponding lifetimes (Fig. \ref{fig9_10_4} d--f) of multicluster amplitude chimeras in the ($P,\sigma$)-plane for different values of time delay: $\tau=0$ (instantaneous coupling); $\tau=\pi/4$; $\tau=\pi$. 
Note that including delay in the coupling results in the appearance of stable multicluster amplitude chimeras and related partially incoherent regimes (white region) lasting longer than the simulation time. The spatio-temporal dynamics of these partially incoherent patterns becomes much richer than in the simple amplitude chimera case, and a few examples of dynamical regimes are provided in Fig. \ref{multicluster_examples}. At the same time, the region of oscillation death changes non-monotonically with the delay, expanding first, but then significantly shrinking in the parameter interval shown, in favor of the synchronized region. 

\subsection{Time-varying delayed coupling}

\begin{figure*}
\centering
\includegraphics[width=0.32\textwidth]{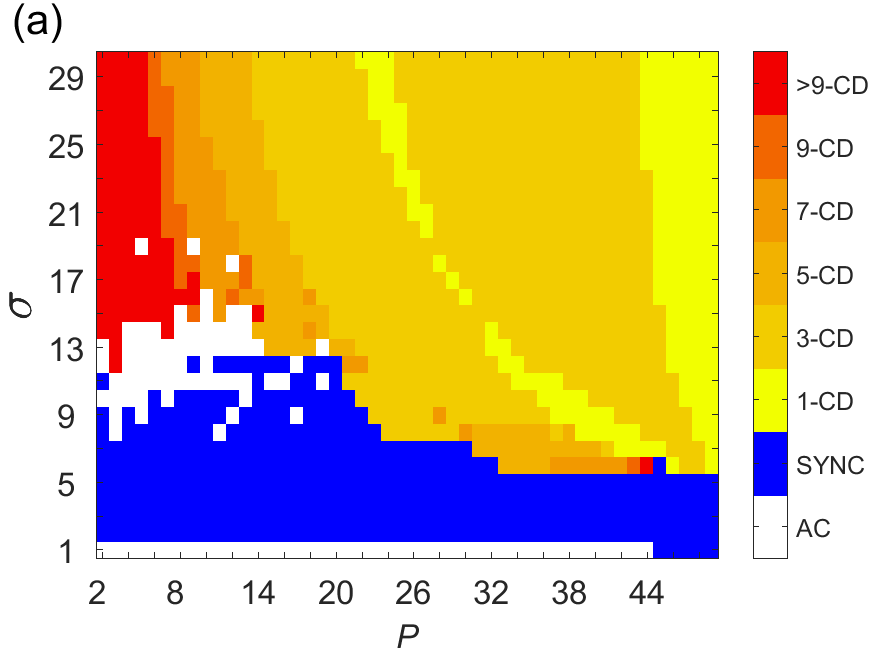}
\includegraphics[width=0.32\textwidth]{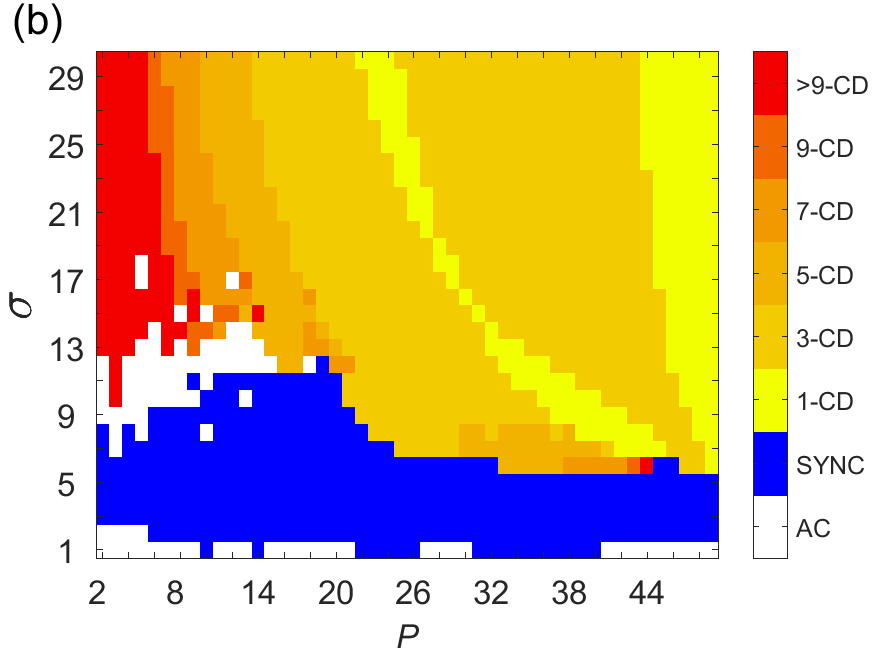}
\includegraphics[width=0.32\textwidth]{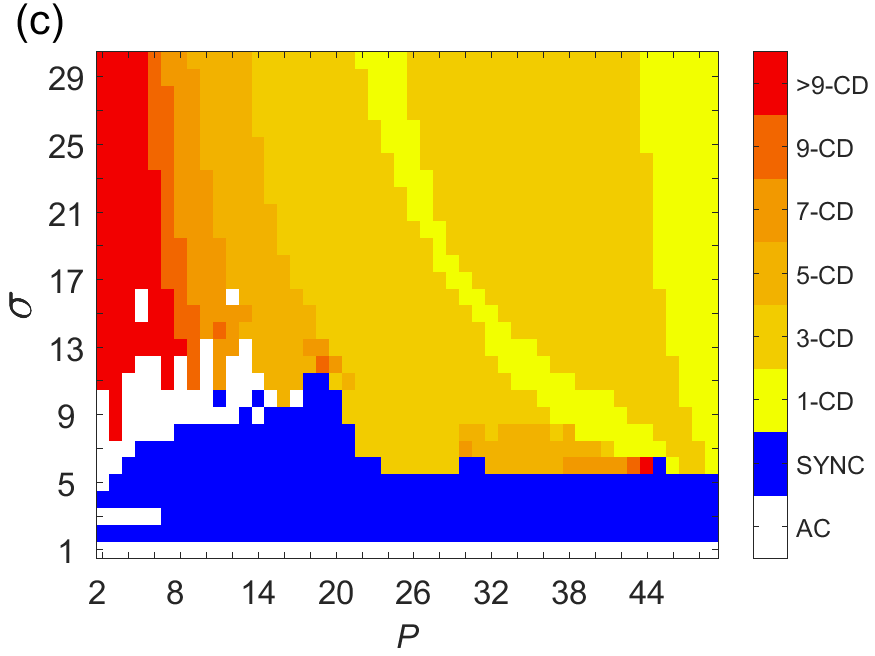}\\
\includegraphics[width=0.32\textwidth]{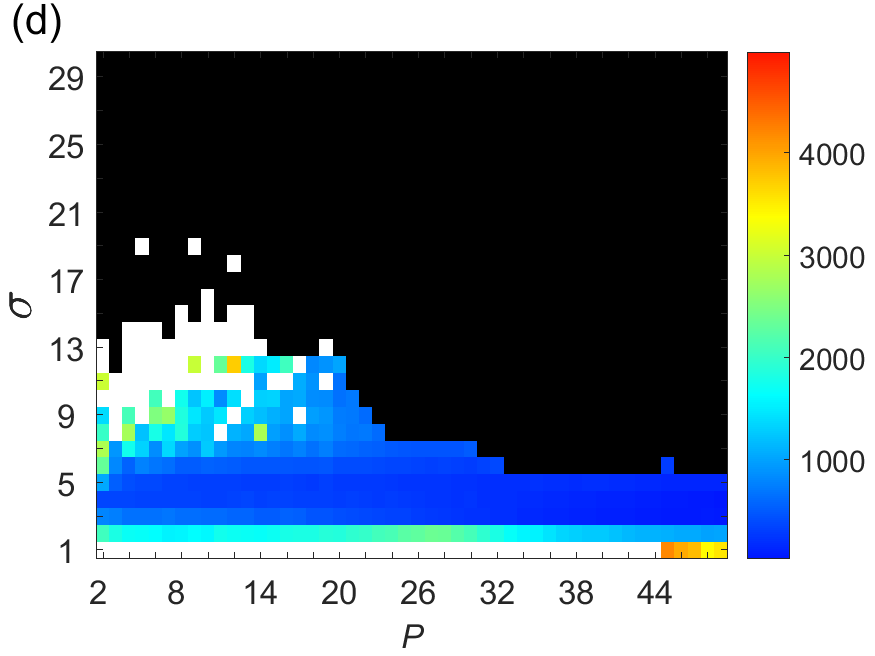}
\includegraphics[width=0.32\textwidth]{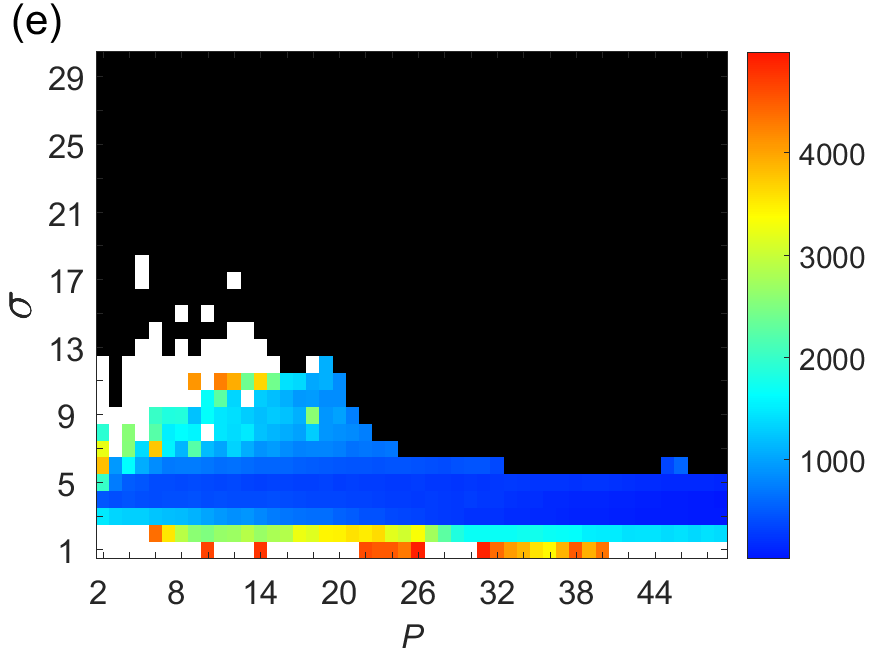}
\includegraphics[width=0.32\textwidth]{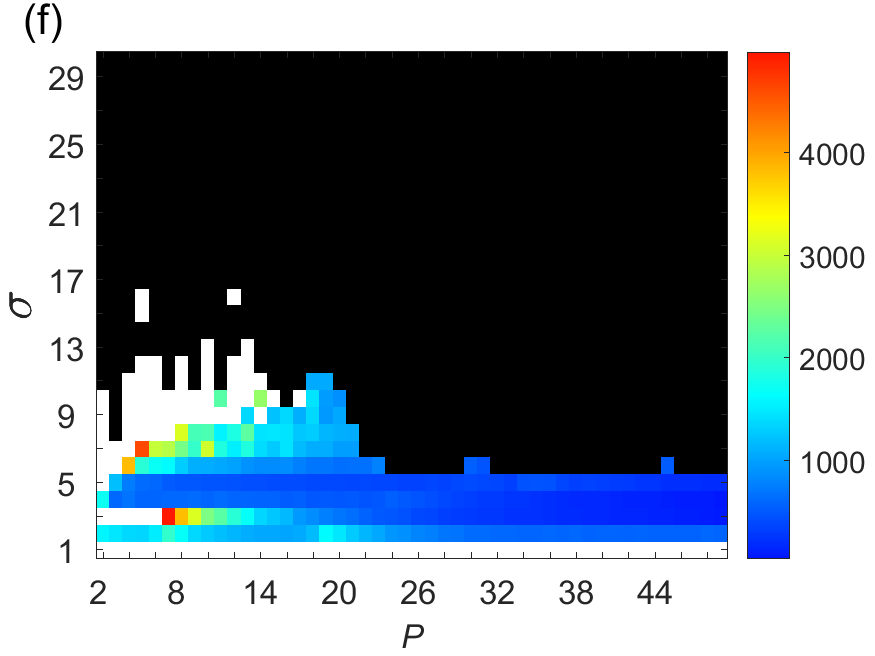}
\caption{(Color online) Map of dynamic regimes (a)--(c) and corresponding chimera lifetimes (d)--(f) in the plane of
coupling range $P$ and coupling strength $\sigma$, for time-varying delay coupling with a sawtooth-wave modulation. 
Nominal time delay $\tau_0=\pi$, modulation frequency $\varpi=10$. 
Modulation amplitude: (a), (d) $\varepsilon = \pi/2$;  (b), (e) $\varepsilon=3\pi/4$; (c), (f) $\varepsilon=\pi$.
Color scale in (a)--(c): 1-CD: 1-cluster chimera death; 3-CD: 3-cluster chimera death; n-CD: n-cluster chimera death. 
SYNC: coherent states (in phase synchronized oscillations, traveling waves, etc.). 
AC: amplitude chimera and related partially incoherent states. 
Color code in transient time diagrams (d)--(f) indicates the time of transition from partially incoherent states 
(amplitude chimera) to coherent states, the white region denotes amplitude chimeras and related partially incoherent states, 
and the black region denotes stable steady states (death states).  
Other parameters: $\lambda= 1$, $\omega = 2$, $N = 100$.}
\label{fig10}
\end{figure*}

\begin{figure*}
\centering
\includegraphics[width=0.32\textwidth]{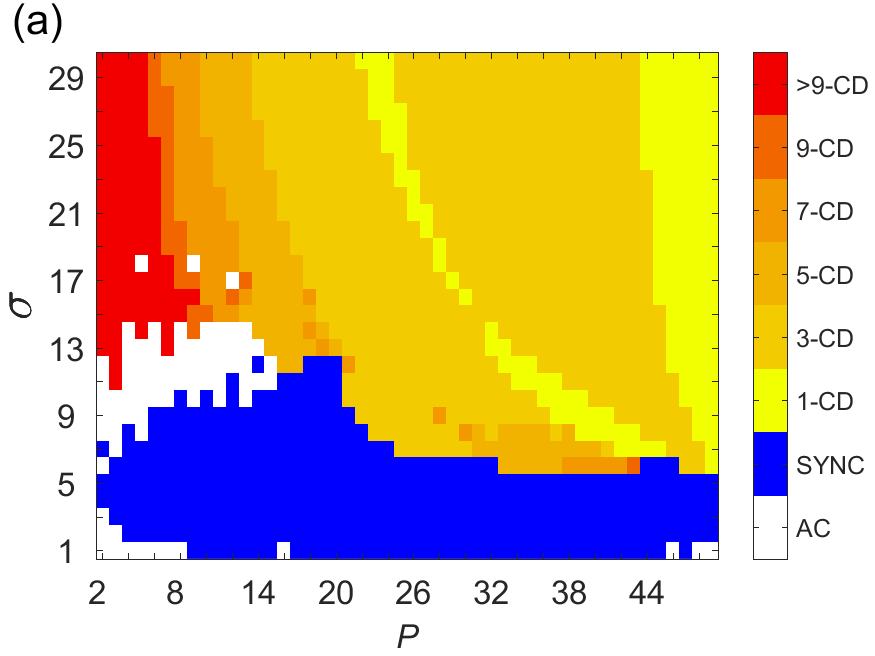}
\includegraphics[width=0.32\textwidth]{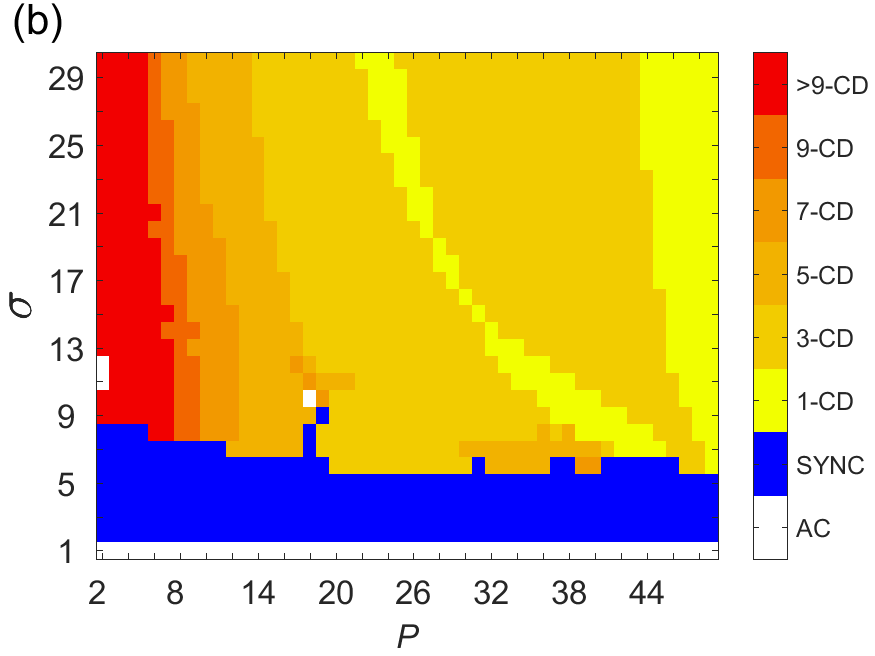}
\includegraphics[width=0.32\textwidth]{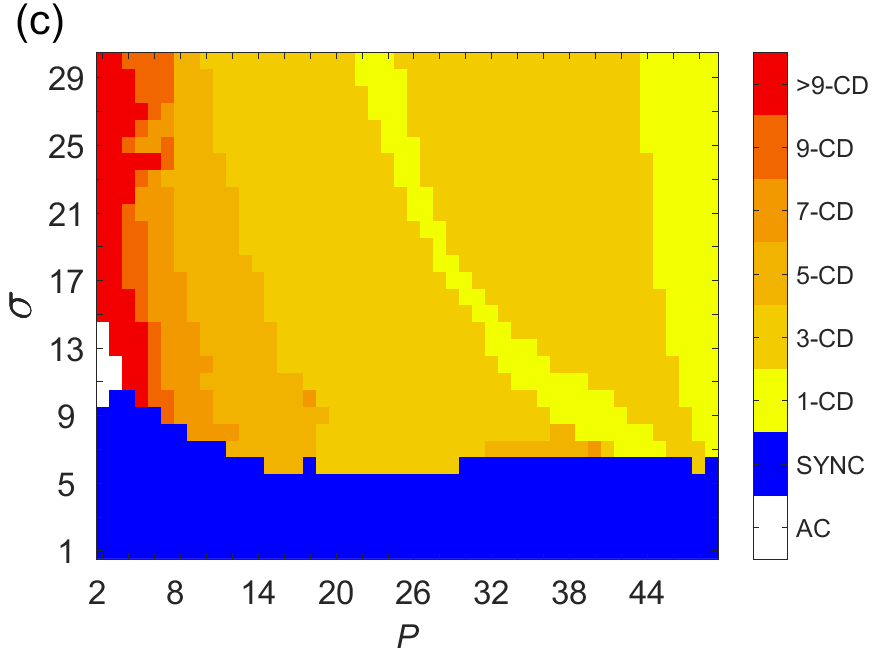}\\
\includegraphics[width=0.32\textwidth]{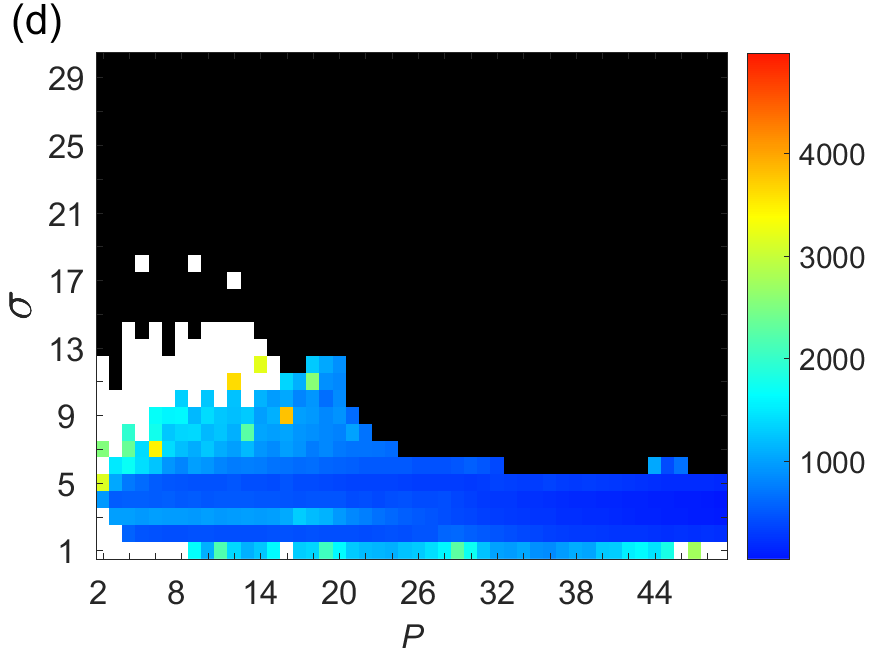}
\includegraphics[width=0.32\textwidth]{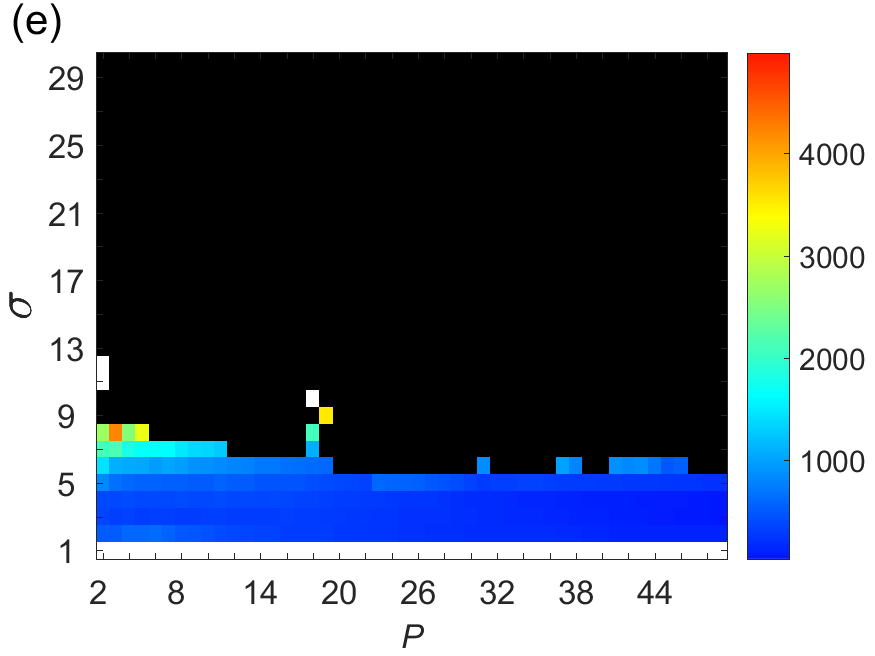}
\includegraphics[width=0.32\textwidth]{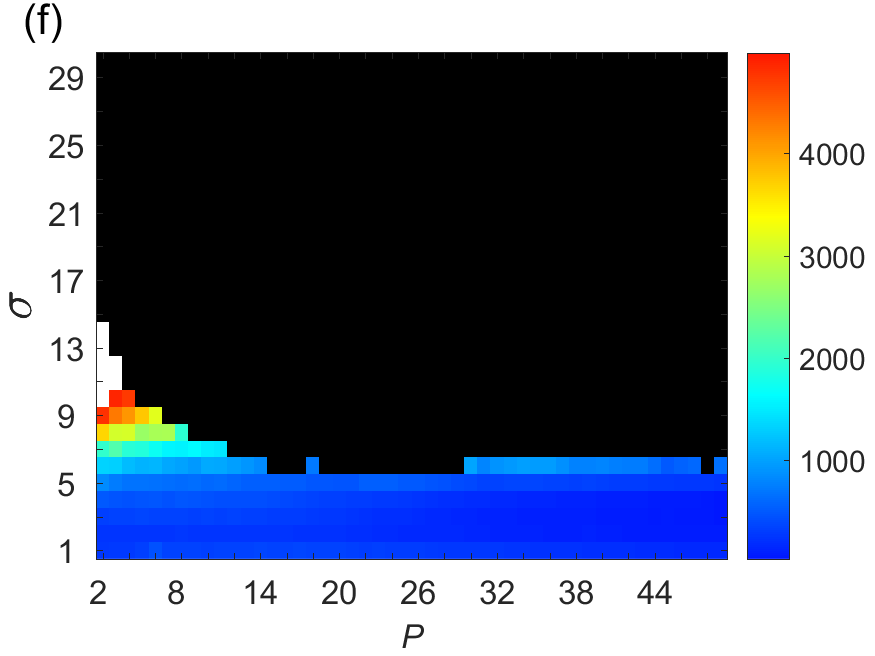}
\caption{(Color online) Same as Fig.~\ref{fig10} for time-varying delay coupling with a square-wave modulation. }
\label{fig11}
\end{figure*}

\begin{figure}
\centering
\includegraphics[width=0.9\columnwidth]{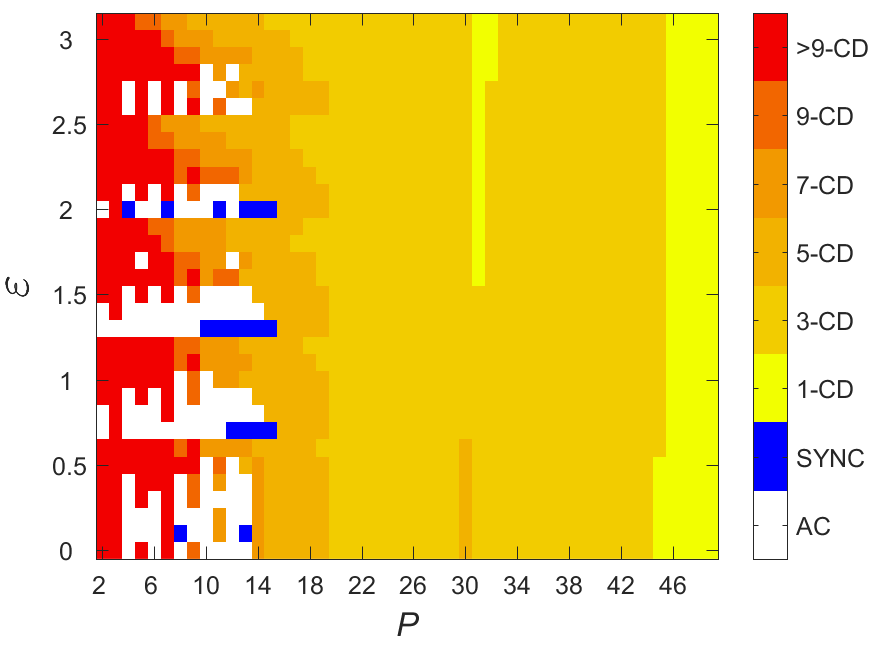}\\
\caption{(Color online) Map of dynamic regimes in the plane of coupling range $P$ and modulation amplitude $\varepsilon$, 
for time-varying delay coupling with a square-wave modulation, and constant value of the coupling strength $\sigma=15$. 
Color scale and other parameters as in Fig.~\ref{fig10}.}
\label{fig12}
\end{figure}

Furthermore, we investigate the influence of time-varying delay on the network dynamics. In particular, we choose a periodic
deterministic modulation of the time-delay around a nominal (average) delay value $\tau_{0}$ in the form of a 
sawtooth-wave modulation \cite{GJU14}:
\begin{align}
\tau(t)&=\tau_{0}+\varepsilon\, \left[2\left(\frac{\varpi t}{2\pi}\, \mathrm{mod}\,\, 1\right)-1\right],
\label{delayvarsawtooth}
\end{align}
and also in the form of a square-wave modulation:
\begin{align}
\tau(t)&=\tau_{0}+\varepsilon\, \mathrm{sgn}[\sin(\varpi t)],
\label{delayvarsquare}
\end{align}
where $\varepsilon$ and $\varpi$ are the amplitude and the angular frequency of the corresponding delay 
modulations, respectively. In particular, for a sawtooth-wave modulation of the delay given by Eq. (\ref{delayvarsawtooth}) we analyze the dynamical states of the network (Fig. \ref{fig10}a--c) and the corresponding lifetimes of partially incoherent states (Fig. \ref{fig10}d--f). We compare the results for different modulation amplitudes: 
$\varepsilon=\pi/2$ (Fig. \ref{fig10}a,d); $\varepsilon=3\pi/4$ (Fig. \ref{fig10}b,e); $\varepsilon=\pi$ (Fig. \ref{fig10}c,f). 
The nominal delay is set to $\tau_0=\pi$, and the angular frequency of the modulation is fixed to $\varpi=10$. 
The initial conditions and the history function are chosen the same as in the constant delay case (Fig. \ref{fig4_5}a,b). The related results for a square-wave modulation of the delay (Eq. \ref{delayvarsquare}) are shown in Fig. \ref{fig11}. 
Note that the sawtooth-wave modulation of the delay does not have a significant influence 
on the various regimes if compared to the constant delay case (Fig. \ref{fig5}). There is, however, 
an occasional appearance of partially incoherent states at small values of coupling strength $\sigma$ and different values of the number of nearest neighbours $P$. They survive the simulation time, but the main region of amplitude chimeras around $\sigma=13$ and small $P$ is mostly unchanged with increasing modulation amplitude (Fig. \ref{fig10}). The square-wave delay modulation is rather interesting, 
since in this case by increasing the modulation amplitude, the domains corresponding to partially incoherent states become drastically reduced, almost disappearing for larger values of modulation amplitude. The impact of the modulation of the coupling delay on the network dynamics in the square-wave case becomes more visible in the parameter plane of the coupling range $P$ and the amplitude of delay modulation $\varepsilon$ for constant coupling strength $\sigma=15$ (Fig. \ref{fig12}). 
One can clearly see that increasing the modulation amplitude $\varepsilon$ results in a sequence of appearance 
and disappearance of the amplitude chimera regions. Such behavior is a characteristic feature of systems  
under square-wave delay modulation, and it has already been reported, for instance, in variable-delay feedback control 
with respect to the sequence of stability islands for successful fixed-point control \cite{GJU13}. 

\subsection{Distributed-delay coupling}

Finally, we consider 
distributed-delay coupling. It has been shown that a time-varying delay system with a high-frequency modulation of the delay time is effectively equivalent to a distributed-delay system with a related delay distribution in the interval of delay variation, and this holds both analytically and numerically with respect to the \emph{steady-state solutions} of the dynamical equations of the delayed system \cite{GJU14,MIC05}. With this in mind, to check if this still holds for other dynamical states of the network, we aim to investigate distributed-delay coupling kernels which correspond to the ones for the time-varying delayed coupling in the previous subsection in the high-frequency limit of the delay modulation. Thus, we choose a uniform distribution kernel:
\begin{align}
G(t')&=\left\{
\begin{array}{cc}
\displaystyle{\frac{1}{2\varepsilon}},& t'\in[\tau_0-\varepsilon,\tau_0+\varepsilon]\\
\\
0, &\text{elsewhere}
\end{array} \right.
\label{uniformkernel}
\end{align}
and a two-peak distribution kernel:
\begin{align}
G(t')=\displaystyle{\frac{\delta(t'-\tau_0+\varepsilon)+\delta(t'-\tau_0-\varepsilon)}{2}}
\label{twodeltakernel}
\end{align}
where $\delta(\cdot)$ denotes the Dirac delta function. In this case $\tau_0$ is the mean time delay of each distribution, and $\varepsilon$ is the distribution width.
They correspond to the average (nominal) delay value and the modulation amplitude, respectively, in the time-varying delay coupling case. The uniform distribution kernel corresponds to the high-frequency limit of a sawtooth-wave modulation of the coupling delay, and a two-peak distribution kernel represents the high-frequency limit of a square-wave modulation of the coupling delay.  In the previous subsection, the delay modulation frequency was chosen as $\varpi=10=5\varpi_0$, where $\varpi_0=2\pi/T=2 $ is the intrinsic angular frequency of the uncoupled system. For those parameter values, the time-varying delay systems can be considered in the high-frequency limit, which is confirmed by our simulations. The resulting diagrams show that the network dynamics with distributed-delay coupling indeed corresponds to the dynamics with time-varying delay coupling with a high-frequency delay modulation. The resulting simulations show excellent matching of the maps of dynamic regimes and chimera lifetimes between the systems with distributed-delay coupling and time-varying delay coupling. Since the dynamic regimes include various synchronous and asynchronous solutions, and combinations of both, we may conclude that approximating the high-frequency time-varying delay system  by a distributed-delay system with a corresponding distribution kernel is quite general, extending well beyond steady state solutions of the complex network dynamics.

\section{Conclusions}
In conclusion, we have investigated a ring network of Stuart-Landau oscillators coupled non-locally and through the real part of the complex variable.
While analyzing various space-time patterns observed in this network, special attention has been payed to the transition from transient amplitude chimera states to phase-lag synchronization (traveling waves) and higher-order coherent structures. Since the Kuramoto order parameter cannot be used as an indicator for such a transition, we have developed a measure which generalizes the global Kuramoto mean-field order parameter and allows us to detect the transition from partially incoherent states (e.g., amplitude chimeras) to any type of coherent structures. 

Further, we have systematically studied the impact of time delay in the coupling on the dynamics of the system, comparing the results for constant, distributed, and time-varying delays with different modulation types.
It has been shown that time delay changes the structure of dynamical regimes of the network in different parameter planes. In particular, time delay induces novel long-living patterns and significantly enhances the lifetime of transient states, specifically amplitude chimeras and related partially incoherent states. Therefore, time delay can be used to control the lifetime of chimera states. Additionally, time delay allows us to construct a desired type of amplitude chimera, for example, with a certain number of clusters or asymmetric cluster configuration.
By appropriately adjusting the modulation of the coupling delay (e.g., square-wave modulation), or equivalently, changing the type and the parameters of the distributed-delay kernel (e.g., two-peak distribution kernel), one can as well reduce the lifetime of amplitude chimeras. Therefore, time delay in the coupling provides a powerful tool to control chimera patterns and their lifetimes in networks of coupled oscillators.

Another result that we found numerically is that at high-frequency delay modulation, the system with time-varying delay coupling is equivalent to distributed delay coupling with related delay-distribution kernels.

\subsection*{Acknowledgment}
This work was supported by DFG in the framework of SFB 910. We thank Sarah Loos, Julien Siebert, and Carolin Wille for helpful discussions.


\end{document}